\documentclass[%
reprint,
 amsmath,assume,
aps,
]{revtex4-2}

\usepackage[dvipdfmx]{graphicx}\usepackage{dcolumn}
\usepackage{dcolumn}
\usepackage{bm}
\usepackage{tabularx}%
\usepackage{color}%
\usepackage{booktabs}%

\newcolumntype{C}{>{\centering\arraybackslash}X}
\newcolumntype{L}{>{\raggedright\arraybackslash}X}
\newcolumntype{R}{>{\raggedleft\arraybackslash}X}

\newcommand{\piY}{\pi Y}
\newcommand{\piYN}{\pi Y \! N}
\newcommand{\piYNN}{\pi Y \! N\!N}

\newcommand{\KN}{\bar{K}\!N}
\newcommand{\KNN}{\bar{K}\!N\!N}
\newcommand{\KNNplus}{\bar{K}\!N\!N^{(I_3=+1/2)}}

\newcommand{\MeV}{{\rm MeV}}

\newcommand{\MeVcc}{{\rm MeV}/c^{2}}

\newcommand{\GeVc}{{\rm GeV}/c}
\newcommand{\GeVcc}{{\rm GeV}/c^{2}}
\newcommand{\MeVee}{{\rm MeV}ee}
\newcommand{\QF}{{\rm QF}_{\bar{K}-{\rm abs} } }

\begin{document}

\preprint{APS/123-QED}

\title{Measurement of the mesonic decay branch of the $\KNN$ quasi-bound state}
  \newcommand{\KEK}{$^1$}
  \newcommand{\RIKEN}{$^2$}
  \newcommand{\RCNP}{$^3$}
  \newcommand{\Victoria}{$^4$}
  \newcommand{\Seoul}{$^5$}
  \newcommand{\IFINHH}{$^6$}
  \newcommand{\SMI}{$^7$}
  \newcommand{\Torino}{$^8$}
  \newcommand{\TorinoU}{$^9$}
  \newcommand{\Frascati}{$^{10}$}
  \newcommand{\Osaka}{$^{11}$}
  \newcommand{\Kyoto}{$^{12}$}
  \newcommand{\Tokyo}{$^{13}$}
  \newcommand{\OsakaE}{$^{14}$}
  \newcommand{\Nishina}{$^{15}$}
  \newcommand{\TITEC}{$^{16}$}
  \newcommand{\TUM}{$^{17}$}
  \newcommand{\komaba}{$^{18}$}
  \newcommand{\Tohoku}{$^{19}$}
  \newcommand{\KIRAMS}{$^{20}$}
  \newcommand{\JAEA}{$^{21}$}
  \newcommand{\Sweden}{$^{22}$}
  \newcommand{\ELPH}{$^{23}$}
  \newcommand{\Chubu}{$^{24}$}
  \newcommand{\CREF}{$^{25}$}

  \author{
    T.~Yamaga\KEK
  }
  \email{takumi.yamaga@kek.jp}

  \author{
    S.~Ajimura\RCNP
  }
  \author{
    H.~Asano\RIKEN
  }
  \author{
    G.~Beer\Victoria
  }
  \thanks{deceased}
  \author{
    H.~Bhang\Seoul
  }
  \author{
    M.~Bragadireanu\IFINHH
  }
  \author{
    P.~Buehler\SMI
  }
  \author{
    L.~Busso\Torino$^,$\TorinoU
  }
  \author{
    M.~Cargnelli\SMI
  }
  \author{
    S.~Choi\Seoul
  }
  \author{
    C.~Curceanu\Frascati
  }
  \author{
    S.~Enomoto\KEK
  }
  \author{
    H.~Fujioka\TITEC
  }
  \author{
    Y.~Fujiwara\Tokyo
  }
  \author{
    T.~Fukuda\OsakaE
  }
  \author{
    C.~Guaraldo\Frascati
  }
  \author{
    T.~Hashimoto\JAEA
  }
  \author{
    R.~S.~Hayano\Tokyo
  }
  \author{
    T.~Hiraiwa\RCNP
  }
  \author{
    M.~Iio\KEK
  }
  \author{
    M.~Iliescu\Frascati
  }
  \author{
    K.~Inoue\RCNP
  }
  \author{
    Y.~Ishiguro\Kyoto
  }
  \author{
    T.~Ishikawa\Tokyo
  }
  \author{
    S.~Ishimoto\KEK
  }
  \author{
    K.~Itahashi\RIKEN$^,$\Nishina
  }
  \author{
    M.~Iwai\KEK
  }
  \author{
    M.~Iwasaki\RIKEN$^,$\Nishina
  }
  \email{masa@riken.jp}
  \author{
    K.~Kanno\Tokyo
  }
  \author{
    K.~Kato\Kyoto
  }
  \author{
    Y.~Kato\RIKEN
  }
  \author{
    S.~Kawasaki\Osaka
  }
  \author{
    P.~Kienle\TUM
  }
  \thanks{deceased}
  \author{
    H.~Kou\TITEC
  }
  \author{
    Y.~Ma\RIKEN$^,$\Nishina
  }
  \author{
    J.~Marton\SMI
  }
  \author{
    Y.~Matsuda\komaba
  }
  \author{
    Y.~Mizoi\OsakaE
  }
  \author{
    O.~Morra\Torino
  }
  \author{
    R.~Murayama\RIKEN
  }
  \author{
    T.~Nagae\Kyoto
  }
  \author{
    H.~Noumi\RCNP$^,$\KEK
  }
  \author{
    H.~Ohnishi\ELPH
  }
  \author{
    S.~Okada\Chubu
  }
  \author{
    H.~Outa\RIKEN$^,$\Nishina
  }
  \author{
    K.~Piscicchia\CREF$^,$\Frascati
  }
  \author{
    Y.~Sada\ELPH
  }
  \author{
    A.~Sakaguchi\Osaka
  }
  \author{
    F.~Sakuma\RIKEN$^,$\Nishina
  }
  \email{sakuma@ribf.riken.jp}
  
  \author{
    M.~Sato\KEK
  }
  \author{
    A.~Scordo\Frascati
  }
  \author{
    M.~Sekimoto\KEK
  }
  \author{
    H.~Shi\SMI
  }
  \author{
    K.~Shirotori\RCNP
  }
  \author{
    D.~Sirghi\Frascati$^,$\IFINHH
  }
  \author{
    F.~Sirghi\Frascati$^,$\IFINHH
  }
  \author{
    S.~Suzuki\KEK
  }
  \author{
    T.~Suzuki\Tokyo
  }
  \author{
    K.~Tanida\JAEA
  }
  \author{
    H.~Tatsuno\Sweden
  }
  \author{
    M.~Tokuda\TITEC
  }
  \author{
    D.~Tomono\RCNP
  }
  \author{
    A.~Toyoda\KEK
  }
  \author{
    K.~Tsukada\Tohoku
  }
  \author{
    O.~Vazquez~Doce\Frascati$^,$\TUM
  }
  \author{
    E.~Widmann\SMI
  }
  \author{
    T.~Yamazaki\Tokyo$^,$\RIKEN
  }
  \author{
    H.~Yim\KIRAMS
  }
  \author{
    Q.~Zhang\RIKEN
  }
  \author{
    J.~Zmeskal\SMI
  }

  \collaboration{J-PARC E15 Collaboration}

  \affiliation{
    \KEK High Energy Accelerator Research Organization (KEK), Tsukuba, 305-0801, Japan
  }
  \affiliation{
    \RIKEN RIKEN Cluster for Pioneering Research, RIKEN, Saitama, 351-0198, Japan 
  }
  \affiliation{
    \RCNP Research Center for Nuclear Physics (RCNP), Osaka University, Osaka, 567-0047, Japan
  }
  \affiliation{
    \Victoria Department of Physics and Astronomy, University of Victoria, Victoria BC V8W 3P6, Canada 
  }
  \affiliation{
    \Seoul Department of Physics, Seoul National University, Seoul, 151-742, South Korea 
  }
  \affiliation{
    \IFINHH National Institute of Physics and Nuclear Engineering - IFIN HH, Romania
  }
  \affiliation{
    \SMI Stefan-Meyer-Institut f\"{u}r subatomare Physik, A-1090 Vienna, Austria
  }
  \affiliation{
    \Torino Istituto Nazionale di Fisica Nucleare (INFN) Sezione di Torino, Torino, Italy
  }
  \affiliation{
    \TorinoU Dipartimento di Fisica Generale, Universita' di Torino, Torino, Italy
  }
  \affiliation{
    \Frascati Laboratori Nazionali di Frascati dell' INFN, I-00044 Frascati, Italy
  }
  \affiliation{
    \Osaka Department of Physics, Osaka University, Osaka, 560-0043, Japan
  }
  \affiliation{
    \Kyoto  Department of Physics, Kyoto University, Kyoto, 606-8502, Japan
  }
  \affiliation{
    \Tokyo  Department of Physics, The University of Tokyo, Tokyo, 113-0033, Japan
  }
  \affiliation{
    \OsakaE Laboratory of Physics, Osaka Electro-Communication University, Osaka, 572-8530, Japan
  }
  \affiliation{
    \Nishina RIKEN Nishina Center, RIKEN, Saitama, 351-0198, Japan 
  }
  \affiliation{
    \TITEC Department of Physics, Tokyo Institute of Technology, Tokyo, 152-8551, Japan
  }
  \affiliation{
    \TUM Technische Universit\"{a}t M\"{u}nchen, D-85748, Garching, Germany
  }
  \affiliation{
    \komaba Graduate School of Arts and Sciences, The University of Tokyo, Tokyo, 153-8902, Japan
  }
  \affiliation{
    \Tohoku Department of Physics, Tohoku University, Sendai, 980-8578, Japan
  }
  \affiliation{
    \KIRAMS Korea Institute of Radiological and Medical Sciences (KIRAMS), Seoul, 139-706, South Korea
  }
  \affiliation{
    \JAEA ASRC, Japan Atomic Energy Agency, Ibaraki 319-1195, Japan 
  }
  \affiliation{
    \Sweden Department of Chemical Physics, Lund University, Lund, 221 00, Sweden
  }
  \affiliation{
    \ELPH Research Center for Electron Photon Science (ELPH), Tohoku University, Sendai, 982-0826, Japan
  }
  \affiliation{
    \Chubu Department of Mathematical and Physical Sciences, Chubu University, Aichi, 487-8501, Japan
  }
  \affiliation{
    \CREF Centro Fermi-Museo Storico della Fisica e Centro studi e ricerche "Enrico Fermi", 000184 Rome, Italy
  }

\date{\today}

\begin{abstract}
We conducted measurements of $K^- + {^3{\rm He}} \to \piYN + N'$ reactions using a $1~\GeVc$  $K^-$-beam, with the objective of understanding the broad decay width of $\KNN$ (approximately twice as broad as that of $\Lambda(1405)$ considered to be the $\KN$ quasi-bound state).
We successfully reproduced distributions of the $\piYN$ invariant mass and momentum transfer for $\piYN$ using model fitting functions for $\KNN$ formation and quasi-free $\bar{K}$ absorption ($\QF$) processes.
The model can describe the experimental data quite well, and four $\KNN \to \piYN $ cross-sections were obtained.
The results indicate that mesonic decay is the dominant decay branch of $\KNN$.
The results also suggest that $\Gamma_{\pi \Lambda N} \sim \Gamma_{\pi \Sigma N}$, which indicates that the $I_{\KN}=1$ absorption channel, in addition to the $I_{\KN}=0$ absorption channel, substantially contribute to the $\KNN$ decay, making the $\KNN$ state approximately twice as unstable as $\Lambda$(1405). 
\end{abstract}

\maketitle


\section{Introduction}
\label{sec:intro}
Quasi-bound systems of an antikaon ($\bar{K}$) and nucleus---the so-called kaonic nuclei---represent an exotic nuclear state in which a meson exists as a constituent particle inside nuclei.
The existence of a kaonic nucleus is predicted as a natural extension of the interpretation of the $\Lambda(1405)$, which is not considered a simple three-quark state but rather a $\bar{K}$-$N$ hadronic molecule attributable to the strongly attractive $\KN$ interaction in the isospin $I_{\KN}=0$ channel.
Neither knowledge of the properties of kaonic nuclei, such as their binding energies, widths, and branching ratios, nor even confirmation of the existence of kaonic nuclei, is essential to construct a precise model of the $\KN$ interaction.
Understanding these properties enables us to determine the energy dependence of the $\KN$ interaction or the $\bar{K}$-absorption strength to nucleon(s) in the energy region below the $\KN$ mass threshold. This energy region cannot be directly probed by $\KN$ scattering or $K^-$-atom experiments.

Among kaonic nuclei, the lightest system, $\KNN$, which is considered to have quantum numbers $I=1/2$ and $J^P = 0^-$, must be understood for the system's fundamental features to be elucidated. 
We conducted the J-PARC~E15 experiment to search for the $\KNN$ and reported observing the $I_3 = +1/2~\KNN$ state, also known as "$K^- pp$," produced in the $^3{\rm He}(K^-,~n)$ reactions and decaying into the non-mesonic $\Lambda p$ decay channel~\cite{exp/jparc/e15/plb2019,exp/jparc/e15/prc2020}.
The binding energy and decay width for $\KNNplus$ were determined to be $B_{\KNN} \sim 40~\MeV$ and $\Gamma_{\KNN} \sim 100~\MeV$, respectively, as the Breit--Wigner parameters.

In theoretical studies, the binding energy and decay width for $\KNN$ have been calculated to be in the range $10-50~\MeV$ and $40-70~\MeV$, respectively, using the chiral-SU(3)-based potential \cite{theor/knucl/dote/plb2018,theor/knucl/barnea/plb2012,theor/knucl/kanada/epja2021,theor/knucl/marri/epja2019,theor/knucl/ohnishi/prc2017,theor/knucl/revai/prc2014,theor/knucl/maeda/pjasb2013}.
In theoretical studies with phenomenological potentials, substantially larger binding energies (by $\sim 20~\MeV$) and similar decay widths have been reported \cite{theor/knucl/kanada/epja2021,theor/knucl/marri/epja2019,theor/knucl/ohnishi/prc2017,theor/knucl/revai/prc2014,theor/knucl/maeda/pjasb2013}. Therefore, the obtained $B_{\KNN}$ is within the range of the theoretical calculations. 
However, the obtained $\Gamma_{\KNN}$ is larger than predicted.
Because the $\Lambda(1405)$ resonance is considered to be the $\KN$ quasi-bound state, the dominant decay branch of $\KNN$ is naively expected to be $\KNN \to \pi \Sigma N$, similar to the $\Lambda(1405) \to \pi \Sigma$ decay (100\%). We would then expect $\Gamma_{\KNN}$ to be similar to the decay width of $\Lambda(1405)$.
However, the obtained $\Gamma_{\KNN}$ is approximately twice as large as $\Gamma_{\Lambda(1405)}$.

Two reasonable scenarios can explain the large $\Gamma_{\KNN}$.
First, the $\KNN$ might decay into non-mesonic $Y \! N$ channels with strength similar to the $\pi \Sigma N$ channel.
This non-mesonic decay occurs via $\bar{K}$ absorption by two nucleons in the nucleus, which most theoretical calculations have not considered.
Second, the $I_{\KN}=1$ interaction might play an important role, similar to the $I_{\KN}=0$ interaction, at least in the absorption channels ({\it i.e.}, $\KNN \to [\piY]_{I=1} N$), in contrast to the theoretical predictions \cite{theor/knucl/sekihara/npa2013,theor/knucl/sekihara/prc2012}.
These two decay modes cannot occur in the decay of $\Lambda(1405)$ and newly open up in the $\KNN$ three-body system.
Therefore, to clarify why $\Gamma_{\KNN}$ is much larger than $\Lambda(1405)$, studying the mesonic decay modes is critical.

To investigate the mesonic decay branches of $\KNN$, we analyzed $K^-+^3{\rm He}$ reactions resulting in $\piYNN$ mesonic final states using data from the J-PARC E15 experiment.
The measured final states in this study are $\pi^- \Lambda p p$, $\pi^+ \Lambda nn$, and $\pi^\mp \Sigma^\pm pn$.
As described in Ref.~\cite{exp/jparc/e15/prc2020}, the $\KNN$ state is produced through the $\bar{K}$ and $N$ substitution via a nucleon knockout reaction on ${^3 \rm He}$. 
Therefore, we focus on the $\piYNN$ channels that originate from the reaction processes
\begin{equation}
\label{eq:knockout}
K^- +N \to {\bar K}+N',
\end{equation}
where $N'$ is emitted in the forward direction. The backscattered intermediate $\bar K$ is subsequently absorbed by the two spectator nucleons as
\begin{equation}
\label{eq:KNN_to_piYN}
{\bar K} + NN \to \piYN,
\end{equation}
where $\KNN$ production could contribute.
The relevant $K^- + {^3 \rm He}$ reaction channels of interest can be summarized as
\begin{equation}
\{\pi^- \Lambda p\}+p' \to [\pi^- (\pi^- p) \,p]_{\rm det} + p_{\rm miss},
\label{eq:1}
\end{equation}
\begin{equation}
\{\pi^+ \Lambda n \} + n' \to [\pi^+ (\pi^- p) \,n]_{\rm det} + n_{\rm miss},
\label{eq:2}
\end{equation}
\begin{equation}
\{\pi^\mp \Sigma^\pm p \}+n' \to [\pi^\mp (\pi^\pm n) \,p]_{\rm det} + n_{\rm miss},
\label{eq:3}
\end{equation}
where $\{\piYN\}$ is the particle set to be investigated and $N'$ is the emitted nucleon in the primary reaction (Eq.~(\ref{eq:knockout})). Each final state is reconstructed as described on the right side of the equations, where $[\pi (\pi N) N]_{\rm det}$ represents detected particles and $N_{\rm miss}$ is an undetected particle identified using the missing mass method.
The $\KNN$ signal is expected to be observed in the $[\piYN]$ invariant-mass spectra for the $I_3=-1/2$ state in Eq.~(\ref{eq:1}), and the $I_3=+1/2$ state in Eqs.~(\ref{eq:2}) and (\ref{eq:3}).
Following the analysis in Ref.~\cite{exp/jparc/e15/prc2020}, we performed a two-dimensional analysis on the $[\piYN]$ invariant mass and the momentum transfer to the $[\piYN]$ system to isolate the signal of $\KNN$.

These reaction processes can couple to $Y^*$ resonances. 
In particular, two $Y^*$ resonances below the $\KN$ mass threshold, $\Lambda(1405)$ and $\Sigma(1385)$, would importantly contribute to the $\bar{K}$ absorption inside $\KNN$. 
Therefore, we analyzed the invariant-mass spectra of the $[\piY]$ subsystem to study the contributions of these $Y^*$ resonances.

This article is organized as follows.
In Sec.~\ref{sec:experiment}, we briefly introduce the J-PARC E15 experiment and describe the essential components of the detector system.
In Sec.~\ref{sec:analysis}, we present the analysis procedure used to select the $\piYNN$ channels.
The obtained spectra are shown in the results in Sec.~\ref{sec:results}.
In Sec.~\ref{sec:fit}, we perform a model fitting using the functions introduced in Ref.~\cite{exp/jparc/e15/prc2020}. 
Specifically, we discuss the $\KNN$ production and the quasi-free $\bar{K}$ absorption processes.
In the fitting procedure, we consider a background originating from the misidentification of $N'$ and $N$ in Eqs.~(\ref{eq:knockout}) and (\ref{eq:KNN_to_piYN}), respectively.
Finally, we compare the cross-sections of $K^-+{^3 \rm He}$ reactions leading to $\piYN+N'$ channels with the non-mesonic $\Lambda p + n$ channel.

\section{Experiment}
\label{sec:experiment}
We briefly overview the experimental setup of J-PARC~E15, focusing on the key apparatus used to analyze the $\piYNN$ channels. The experimental configuration has been described in detail in the literature~\cite{exp/jparc/e15/ptep2012-a,exp/jparc/e15/ptep2012-b,exp/jparc/e15/nima2012}. We took the current dataset in parallel to that used for the $K^- + \,^3{\rm He} \to \Lambda p + n$ analysis~\cite{exp/jparc/e15/prc2020} with the same trigger condition. In the experiment, a focused $K^-$ beam interacted with a liquid $^3$He target. The particles produced in the $K^- + {^3 \rm He}$ reaction were detected by a cylindrical detector system (CDS) surrounding the target.
The CDS consists of a cylindrical drift chamber (CDC) and a cylindrical hodoscope counter made of a plastic scintillator (CDH).

The CDS operates in a magnetic field of $0.7~{\rm T}$ along the beam axis ($z$-axis). The tracking information provided by the CDC enables the analysis of charged particle momenta. We determined the starting point of the charged particle or the reaction point from the CDC track and $K^-$ beam track by searching for the closest point of the two tracks. In addition, particle identification for each track is performed using momentum and time-of-flight information obtained from the CDH.

For neutron detection, we used the CDH, which consists of plastic scintillators measuring 30~mm in thickness and 790~mm in length. The inner surface of the CDH was positioned at a radius of 544~mm from the beam axis.
Using a Monte Carlo simulation based on Geant4, we estimated the typical neutron detection efficiency to be 3--9\% (at neutron momenta of 1--0.2~$\GeVc$). The neutron momentum was determined using the time-of-flight information and neutron flight length obtained from the reaction point, and the $z$-position along the CDH deduced by the time difference between two PMTs at both ends.

\section{Analysis}
\label{sec:analysis}

We focused on the following four channels: ${ \pi^- \Lambda p } +p'$ (Eq.~(\ref{eq:1})), ${ \pi^+ \Lambda n } + n'$ (Eq.~(\ref{eq:2})), and ${ \pi^\mp \Sigma^\pm p } + n'$ (Eq.~(\ref{eq:3})). Among these four channels, only the ${ \pi^- \Lambda p } + p'$ channel can be measured without detecting a neutron; specifically, we required the detection of $(pp\pi^-\pi^-)$ in the CDS. In the other three channels, neutron detection is necessary, along with detecting the remaining three charged particles ($p \pi^+ \pi^-$).

The CDS detects the ${ \piYN }$ particles in each final state. The undetected particle $N'$ is identified through the missing mass method using the measured $K^-$ beam and the information from the ${ \piYN }$ particles. Other ${ \piYN } + N'$ channels, including multi-neutrons or $\gamma$ in the final decay products, were omitted from the present analysis because the final states cannot be identified unambiguously.


\subsection{Neutron analysis}
We applied the following criteria to identify neutrons. First, we required no CDC track on the fired CDH segment and both sides to eliminate fake neutron signals from charged particles impacting the material close to the CDH segment. Second, we required the energy deposited by the CDH-hit to be greater than $2~\MeVee$ (electron-equivalent) to remove the accidental background recorded in the CDH. Finally, $1/\beta$ was calculated for the CDH-hit using the time-of-flight information, and events with $1/\beta<1.5$ were rejected to eliminate $\gamma$-ray background originating from the $K^- + {{^3}{\rm He}}$ reaction, which peaks at $1/\beta \approx 1$.
 
\begin{figure}[ht]
    \centering
    \includegraphics[width=\linewidth]{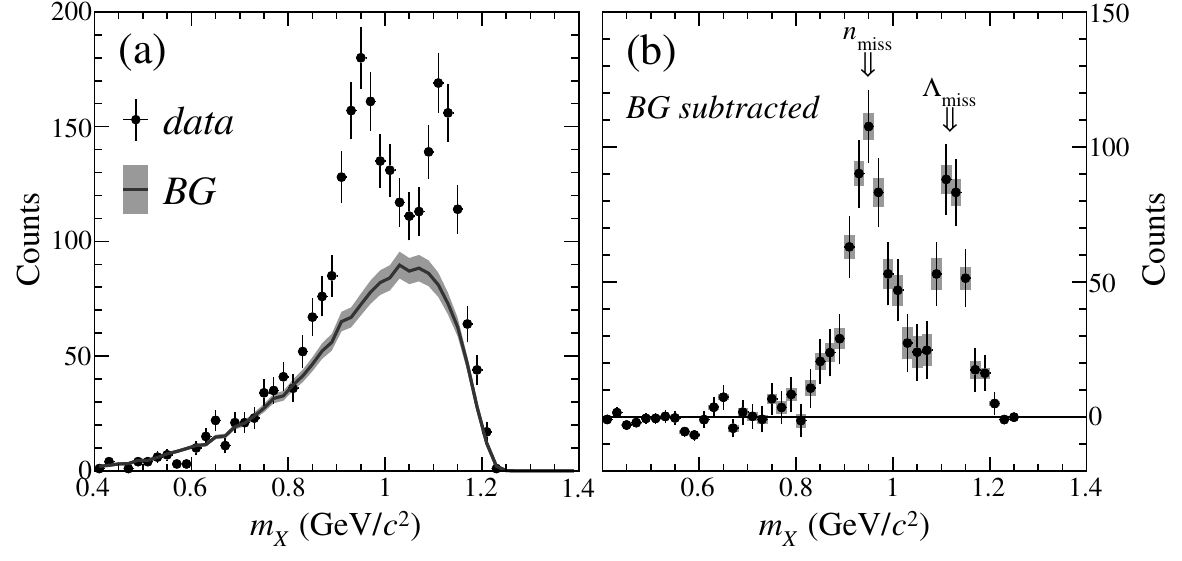}
    \caption{
        Mass distributions of missing particle $X$ in the ($pn\pi^+\pi^- + X$) event selection: (a) distribution before background subtraction and (b) distribution after background subtraction.
    }
    \label{fig:fake_neutron}
\end{figure}
 
Even after implementing the aforementioned background rejection procedures, our data still contain numerous neutron-like fake signals in the neutron's missing mass spectrum. Figure~\ref{fig:fake_neutron}(a) illustrates the distribution of missing mass ($m_X$) in the $K^- + {{^3}{\rm He}} \to pn\pi^+ \pi^- + X$ events. 
The figure shows that the signal-to-noise ratio is approximately 1:1, necessitating a thorough evaluation of the background effect to achieve a detailed analysis.
 
The fake neutron signals are generated through multiple reactions or scattering, reach the CDH, and are identified as neutral particles.
We therefore evaluated the effect of these fake signals by simulating random CDH hits using an event-mixing method.
In this method, we disconnected the CDH-hit information from other charged-particle-hit information and randomly recombined the two datasets.
The gray hatched curve in the figure represents the result obtained when this event-mixing method was applied.
We determined a scaling factor for the background distribution by fitting the $m_{X}$ distribution below $0.8~\GeVcc$, where the data should only include neutron contamination events.

Figure~\ref{fig:fake_neutron}(b) shows that the background subtraction was successful in the $m_X=m_n$ region and that the $[pn\pi^+ \pi^-] + n_{\rm miss}$ final state clearly remains as a peak in the missing-mass spectrum.
This background subtraction procedure also agrees in the higher-mass region up to 1.2 GeV/$c^2$, including the $m_X \sim m_\Lambda$ region.
In fact, $\Lambda$ missing events, specifically $[pn \pi^+ \pi^-] +\Lambda_{\rm miss}$ events, also remain clearly visible.
This agreement demonstrates the validity of the fake-neutron subtraction procedure.
We evaluated the fitting error of the scaling factor, as indicated by the gray box in the figure, and considered this error to be a systematic error in the subsequent analysis.

\subsection{Selection of final state}
\label{sec:analysis:event-selection}
For all the selected events, $[pp\pi^- \pi^-] + p_{\rm miss}$ and $[pn\pi^+ \pi^-] + n_{\rm miss}$, we applied a common procedure to identify the final states: $[\pi^- \Lambda p] + p_{\rm miss}$ (Eq.~(\ref{eq:1})), $[\pi^+ \Lambda n] + n_{\rm miss}$ (Eq.~(\ref{eq:2})), and $[\pi^\mp \Sigma^\pm p] + n_{\rm miss}$ (Eq.~(\ref{eq:3})). We used a log-likelihood method, which was introduced in Ref.~\cite{exp/jparc/e15/prc2020}, to determine whether the event kinematics and topology are consistent with the reactions described in Eqs.~(\ref{eq:1}), (\ref{eq:2}), and (\ref{eq:3}).
To assess the kinematic information, we conducted a kinematic fit to evaluate how well the event kinematics matched the final state of interest within the detector's resolution.
For geometrical information, we used the distances of the closest approach (DCA) between two charged particles. These distances should be zero within the resolution of the detector.

We defined the log-likelihood $L$ using probability density functions, $p(x)$, for the $\chi^2$ of the kinematical fit and DCAs, as follows:
\begin{equation}
L = - \ln \left( p(\chi^2_{\text{kfit}}) \times \prod_{i}^{N_{\text{DCA}}} p(\text{DCA}_i) \right),
\end{equation}
where $N_{\text{DCA}}$ is the number of DCAs considered and $p(\chi^2_{\text{kfit}})$ represents the probability function for $\chi^2$ with ${\text{NDF}}_{\text{kfit}}= 2$. 
Function $p(\text{DCA})$ is a Gaussian function with mean $\mu=0$ and standard deviation $\sigma$, evaluated through Monte Carlo simulation with realistic detector resolutions. 
All of the $p(x)$ functions are normalized such that their maximum values are equal to one, making $L=0$ at the most probable point.

\begin{figure}[ht]
    \centering
    \includegraphics[width=\linewidth]{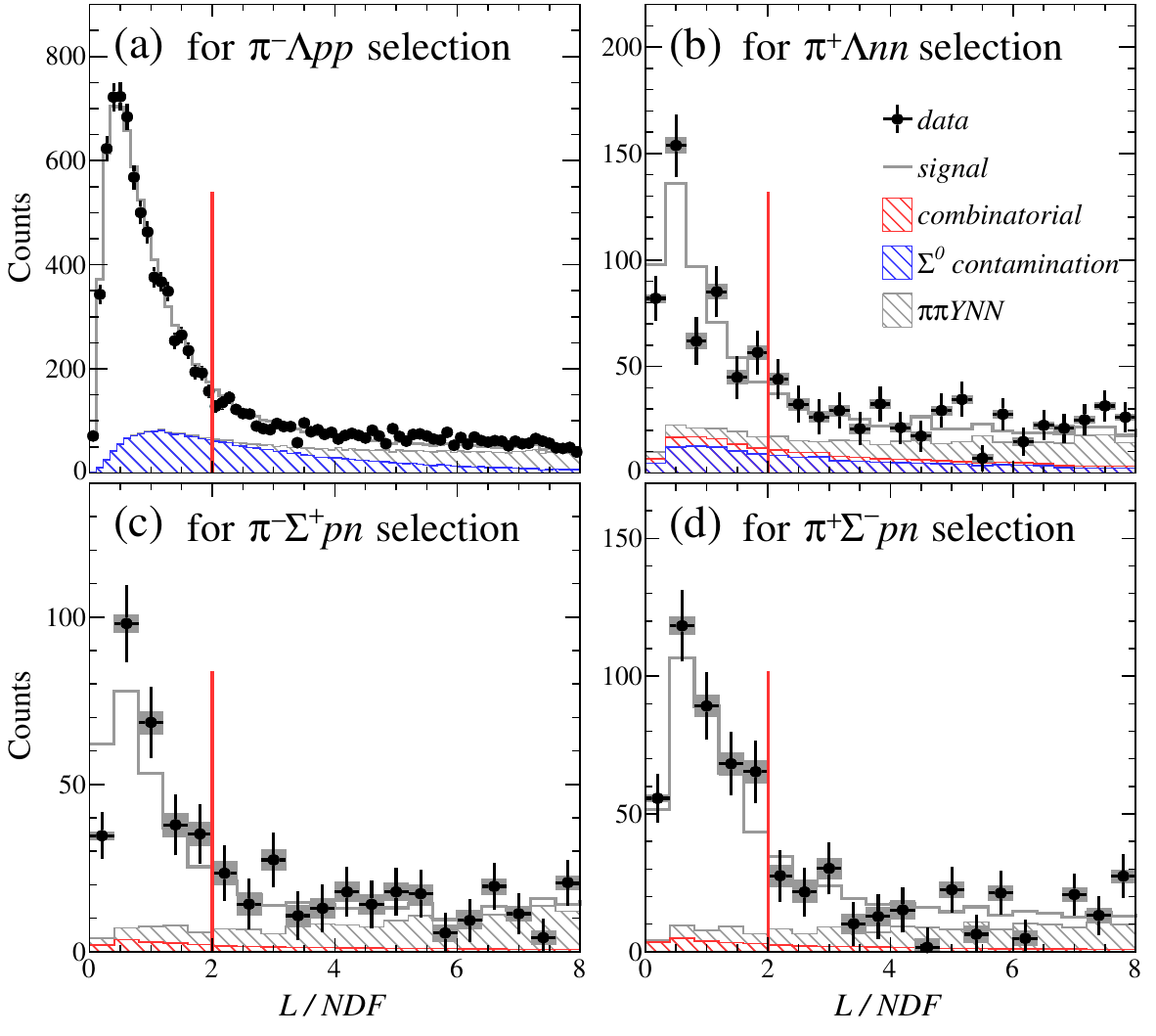}
    \caption{
Log-likelihood distributions for the event selections of (a) $\pi^- \Lambda pp$, (b) $\pi^+ \Lambda nn$, (c) $\pi^- \Sigma^+ pn$, and (d) $\pi^+ \Sigma^- pn$. The red vertical line indicates the threshold $L/{\rm NDF} < 2$, and events falling below this line are selected as the final event samples.
    }
    \label{fig:lnl}
\end{figure}

Multiple-particle combinations could be generated by hyperon decays or $\bar{K}^0 \to \pi^+ \pi^-$ decay. 
We examined all $L$ with possible particle combinations for each event to isolate which particle pairs are the decay products. 
We then determined that the reaction process of the events resulted in the minimum $L$.
Figure~\ref{fig:lnl} illustrates the distributions of the $L/{\rm NDF}$ for each identified process. 
As shown in the figure, the $L/{\rm NDF}$ distributions are quite similar, with the most frequent $L/{\rm NDF}$ values clustering around $L/{\rm NDF} \sim 0.5$. 
The yields decrease as $L/{\rm NDF}$ becomes larger.
To select the final event sets, we accepted events with $L/{\rm NDF}<2$; the threshold is indicated by the vertical red line in the figure.

Because of the finite resolution, the correct particle combination for the reaction process is not always associated with the minimum $L$.
There are three significant sources of contamination: I) combinatorial effects (improper particle pairing in hyperon decay, including $K^0$ contamination), II) $\Sigma^0$ contamination (resulting from a missing $\gamma$ in the $\Sigma^0$ decay), and III) contamination arising from the $\pi \piYNN$ final states.
We assessed the contamination yields in each channel through simulation, as shown in the hatched region in the figure.
The background yields relative to the total events in the $[\pi^- \Lambda p] + p_{\rm miss}$, $[\pi^+ \Lambda n] + n_{\rm miss}$, and $[\pi^\mp \Sigma^\pm p] + n_{\rm miss}$ event sets were found to be $({\rm I}+{\rm II}+{\rm III}) \sim (0+15+0)\%$, $(4+12+6)\%$, and $(5+0+6)\%$, respectively.

\begin{figure}[ht]
    \centering
    \includegraphics[width=\linewidth]{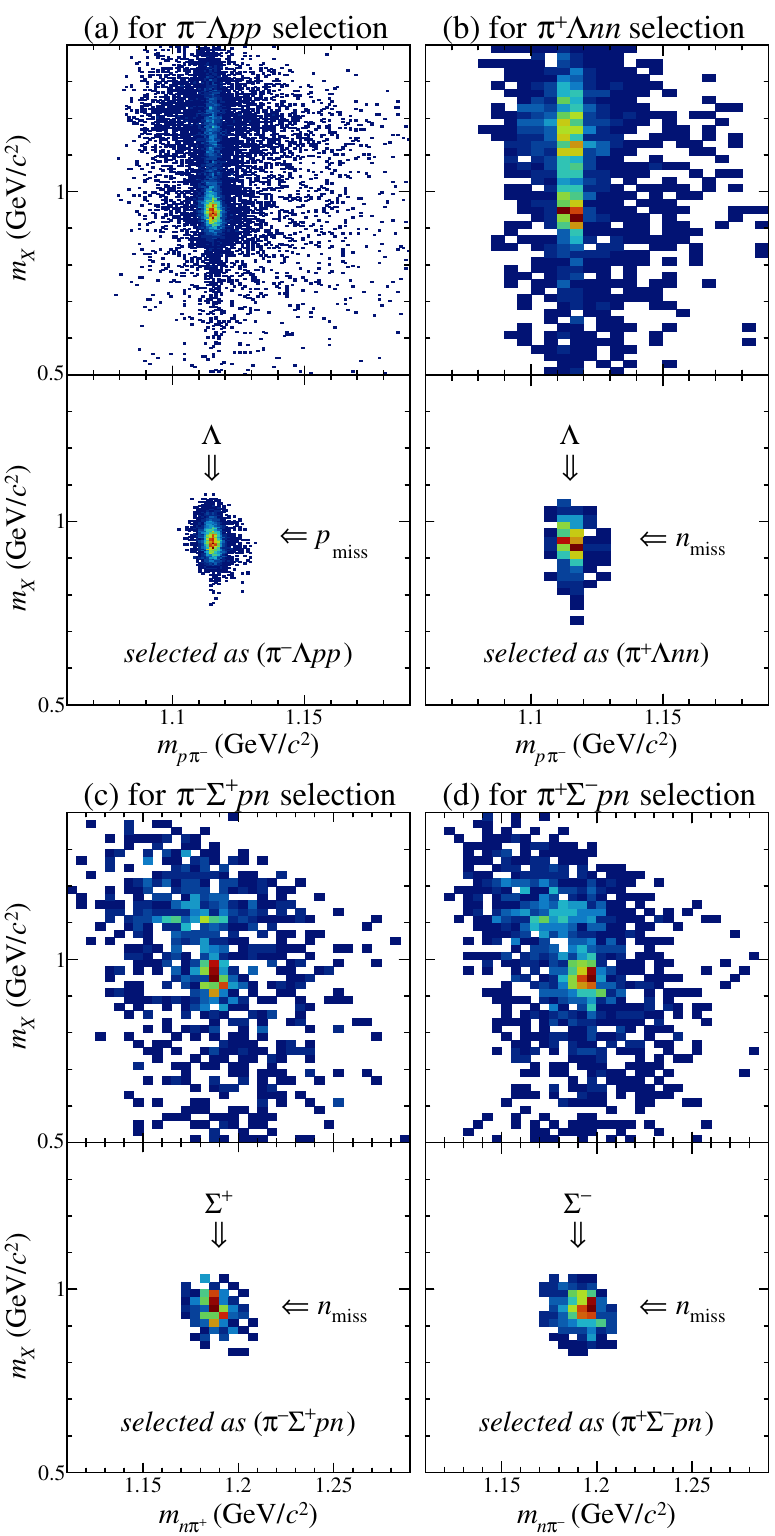}
    \caption{
  Two-dimensional distributions of masses of hyperon candidate ($m_{N\pi}$) and missing particle ($m_{X}$). Panels (a), (b), (c), and (d) correspond to the event selections of $\pi^- \Lambda pp$, $\pi^+ \Lambda nn$, $\pi^- \Sigma^+ pn$, and $\pi^+ \Sigma^- pn$, respectively. The top figures display all events with $L<100$, whereas the bottom figures show the selected events with $L<2~{\rm NDF}$.
    }
    \label{fig:mass_selection}
\end{figure}

To demonstrate the validity of the event selection, we present the event distributions of the invariant mass of the selected $\pi N$ pairs from the hyperon decay ($m_{\pi N}$) and the missing mass of $\piYN$ ($m_X$) in Fig.~\ref{fig:mass_selection}.
The top panel of the figure pair displays events with $L<100$, whereas the bottom panel represents events after the selection.
The signals corresponding to each $[\piYN]+N_{\rm miss}$ final state are clearly observed as concentrated events around $(m_X,~m_{\pi N}) \sim (m_N,~m_Y)$, where $m_N$ and $m_Y$ represent the intrinsic masses of the nucleon and $\Lambda$ or $\Sigma$, respectively.

\subsection{Acceptance correction}
\begin{figure}[ht]
    \centering
    \includegraphics[width=\linewidth]{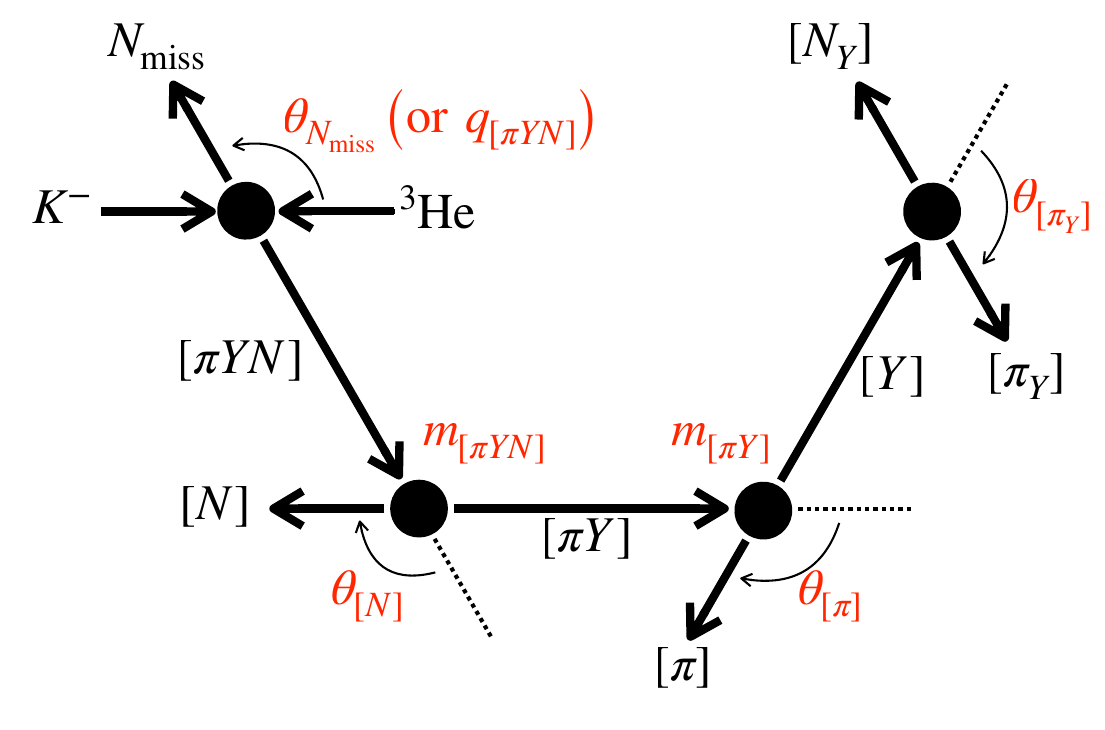}
    \caption{
  Kinematics considered for $K^- + {^3{\rm He}} \to [\piYN] + N_{\rm miss}$ reaction. The figure illustrates the six kinematical parameters (indicated by red) considered for acceptance correction.
    }
    \label{fig:kinematics}
\end{figure}

In the measurement, the acceptance of an event is determined by the four particles to be measured, $[\piYN] \to [\pi(\pi_Y N_Y)N]$.
We consider the event kinematics as a cascading two-body decay scheme as illustrated in Fig.~\ref{fig:kinematics}.
In the reaction process of interest, the strange quark is carried by an intermediate $\bar{K}$, as discussed in Sec.~\ref{sec:intro}. 
Thus, the following equations can describe the process:
\begin{equation}
\begin{split}
\label{eq:b1}
K^- + ~^3{\rm He} \to &[\KNN] + N_{\rm miss}, \\
&[\KNN] \to [\bar{K}N] + [N].
\end{split}
\end{equation}
This reaction corresponds to the two-body decays:
\begin{equation}
\begin{split}
K^- + ~^3{\rm He} \to &[\piYN] + N_{\rm miss}, \\
&[\piYN] \to [\piY] + [N],
\end{split}
\end{equation}
as shown in the figure. The kinematics of each vertex point are defined by three parameters: the polar angle ($\theta$) and azimuthal angle ($\phi$) of the daughters in the rest frame of the parent system, and the parent mass. 
The masses of the initial $K^- + ~^3{\rm He}$ system and the final $\pi_Y + N_Y$ system are constrained to be $\sqrt{s}$ and $m_Y$, respectively. 
Therefore, there are ten independent kinematical parameters: the masses of $[\piYN]$ and $[\piY]$; $\theta$ and $\phi$ of $N_{\rm miss}$; $N$; $\pi$; and $\pi_Y$.

Because the azimuthal angles ($\phi$) do not affect the kinematics of the successive process, we can accurately evaluate the acceptance by integrating $\phi$ under the assumption of a uniform distribution. 
However, the polar angles ($\theta$) cannot be integrated.
Consequently, we must consider the six kinematic parameters in the figure for the acceptance correction.

\begin{figure}[ht]
    \centering
    \includegraphics[width=\linewidth]{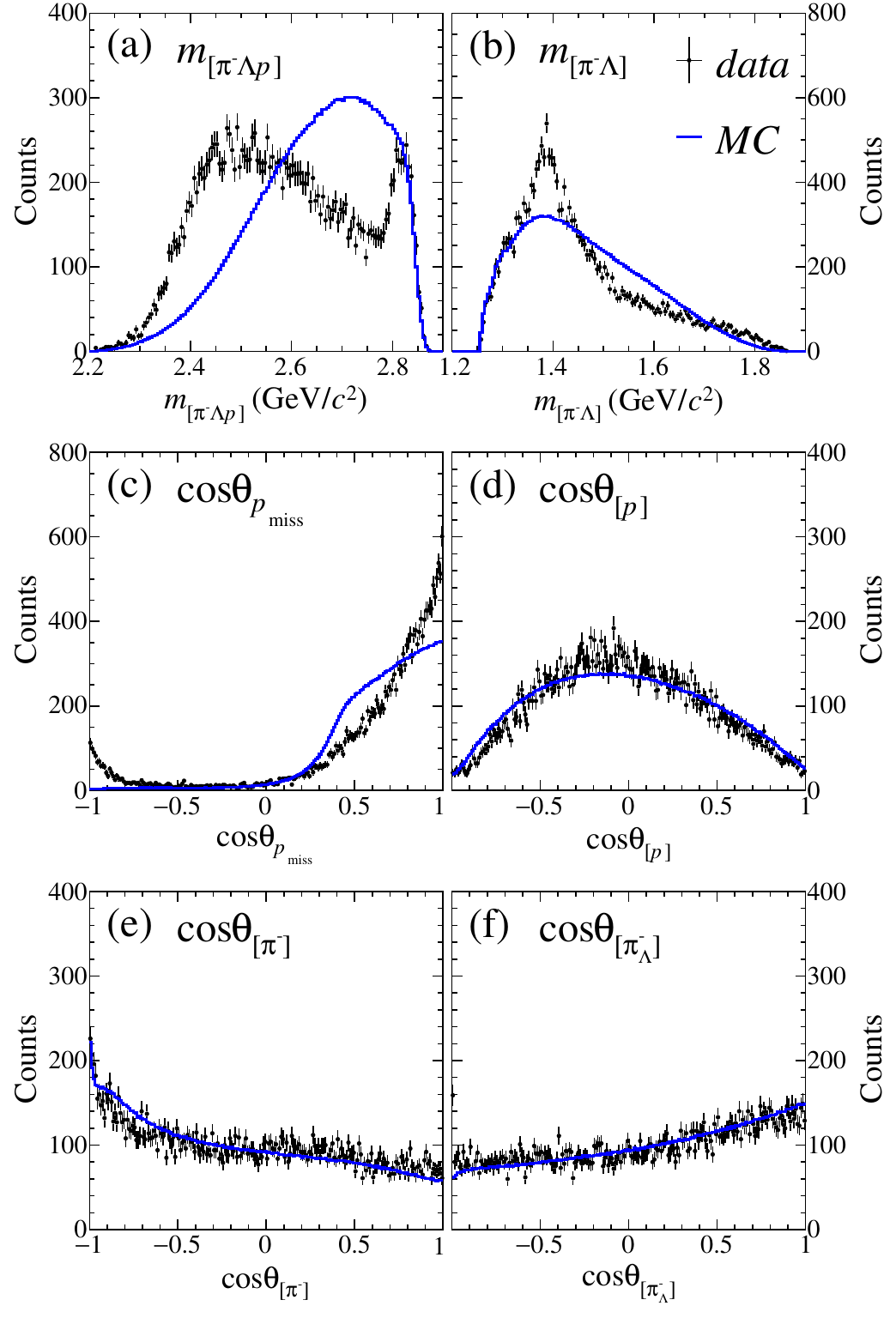}
    \caption{
Distributions of the five kinematical parameters considered in the acceptance correction for the $[\pi^- \Lambda p] + p_{\rm miss}$ channel. The black and blue distributions represent the data and the four-body phase space distribution, including the acceptance. The simulated distributions were scaled to have the same integral as the data.
    }
    \label{fig:dist_for_corr}
\end{figure}

We conducted a Monte Carlo simulation using Geant4 to obtain the acceptance map.
In the simulation, the $K^- + {^3 \rm He} \to \piYNN$ reaction was generated according to the four-body phase space distribution.
Figure~\ref{fig:dist_for_corr} displays the distributions of the five kinematical parameters mentioned in Fig.~\ref{fig:kinematics} for both the experimental data and the simulation.
If the reaction is governed by the phase space only, without any additional physics effects beyond the density of states of the final state, we would expect the data (black) and the simulation (blue) to coincide.

As shown in Fig.~\ref{fig:dist_for_corr}(a)--(c), the data deviate from the simulation; however, good agreement is observed in Fig.~\ref{fig:dist_for_corr}(d)--(f).
Hence, we applied the acceptance correction on an event-by-event basis for these three parameters while integrating and correcting the remaining three using the average acceptance.
In addition, we used $q_{[\piYN]}$ instead of $\theta_{N_{\rm miss}}$ because our interest lies in the momentum transfer to the $[\piYN]$ system rather than in the specific formation angle $\theta_{N_{\rm miss}}$.
Consequently, we constructed a three-dimensional acceptance map, $\varepsilon(m_{[\piYN]},~q_{[\piYN]},~m_{[\piY]})$, which encompasses the geometric acceptance of the CDS, detection efficiencies, and analysis efficiencies.

To investigate the production of $\KNN$ or $Y^*$ resonances, we are specifically interested in three distributions---specifically, $m_{[\piYN]}$, $q_{[\piYN]}$, and $m_{[\piY]}$---for each $[\piYN]+N_{\rm miss}$ channel.
To obtain the differential cross-sections of these three parameters, we initially calculated the triple differential cross-section $d^3 \sigma / dm \, dq \, dm_{[\piY]}$, where $m$ represents $m_{[\piYN]}$ and $q$ represents $q_{[\piYN]}$. The expression is given by
\begin{equation}
\frac{d^3 \sigma}{dm \, dq \, dm_{[\piY]}} = \frac{N(m,\,q,\,m_{[\piY]})-N_{\rm fake}(m, q, m_{[\piY]})}{\mathcal{L} \, \Delta m \, \Delta q \, \Delta m_{[\piY]} \,\varepsilon(m, q, m_{[\piY]})},
\end{equation}
where $\mathcal{L}=2.89\pm0.01~{\rm nb}^{-1}$ represents the integrated luminosity and $N(m,\,q,\,m_{[\piY]})$ is the number of events in the ($m,\,q,\,m_{[\piY]}$) bin with bin widths $\Delta m$, $\Delta q$, and $\Delta m_{[\piY]}$, respectively.
Parameter $N_{\rm fake}(m,\,q,\,m_{[\piY]})$ denotes the expected fake-neutron background yield in the real data, as described in Sec.~\ref{sec:analysis:event-selection} (note that $N_{\rm fake}(m,\,q,\,m_{[\piY]})=0$ for the $[\pi^- \Lambda p] + p_{\rm miss}$ channel because this channel contains no neutrons).

We omitted the low-efficiency region $\varepsilon(m,q,m_{[\piY]}) < 0.1\%$  for the $[\pi^- \Lambda p] + p_{\rm miss}$ and the region $\varepsilon(m,q,m_{[\piY]}) < 0.015\%$ for the other channels.
Finally, we obtained differential cross-sections of interest by integrating the triple-differential cross-sections.

\section{Results}
\label{sec:results}
\begin{figure*}[ht]
    \centering
    \includegraphics[width=\linewidth]{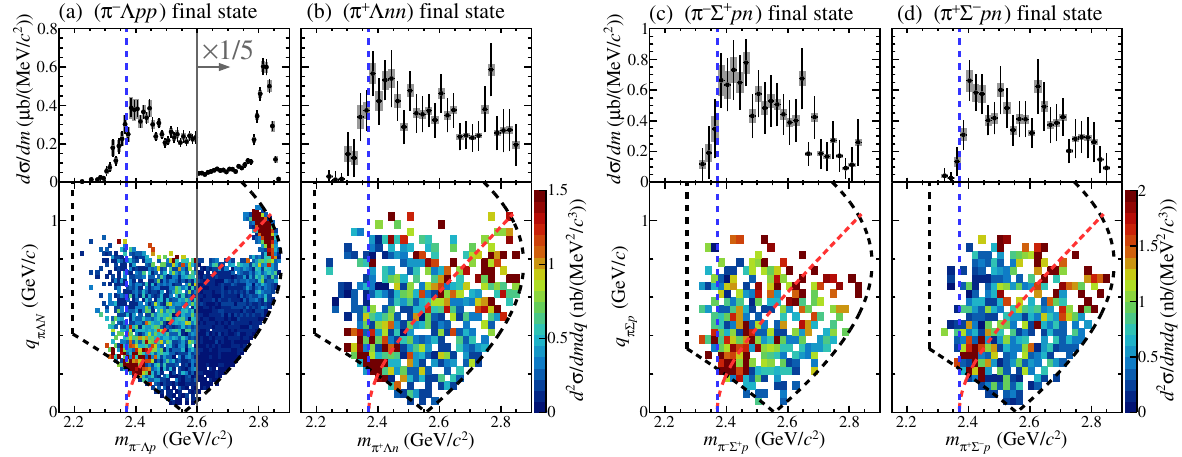}
    \caption{
    Acceptance-corrected $(m_{[\piYN]},q_{[\piYN]})$ distributions and their projections on the $m_{[\piYN]}$ axes are shown in the figure:
(a) corresponds to $[\pi^- \Lambda p] + p_{\rm miss}$,
(b) corresponds to $[\pi^+ \Lambda n] + n_{\rm miss}$,
(c) corresponds to $[\pi^- \Sigma^+ p]+n_{\rm miss}$, and
(d) corresponds to $[\pi^+ \Sigma^- p]+n_{\rm miss}$.
The event concentrations along a kinematical line (Eq.~(\ref{eq:m_qf})) are clearly shown, especially in (a).
For the mass region $m_{\pi^- \Lambda p}>2.6~\GeVcc$, the cross-section in (a) is scaled by a factor of $1/5$ to show the events-concentration corresponding to Eq.~(\ref{eq:2na_pimLp}).
The black dotted lines represent the kinematical limit of the reaction.
The blue vertical dotted lines indicate the mass threshold of $m_{\bar{K}} + 2m_N$.
The red dotted curves represent the quasi-free kinematical line described in Eq.~(\ref{eq:m_qf}).
    }
    \label{fig:m_vs_q_cs}
\end{figure*}

\subsection{($m_{\piYN},~q_{\piYN})$ distributions}
\label{sec:results:m_vs_q}
We conducted measurements of the two-dimensional $(m_{[\piYN]},q_{[\piYN]})$ distributions to study the production of $\KNN$ decaying into mesonic $[\piYN]$ channels.
Figure~\ref{fig:m_vs_q_cs} shows the obtained $(m_{[\piYN]},~q_{[\piYN]})$ distributions for each final state, along with projections onto the $m_{[\piYN]}$ axis.

An event concentration at $(m_{[\piYN]},~q_{[\piYN]}) \sim (2.4~\GeVcc,~0.3~\GeVc)$ was observed consistently in all of the $\piYNN$ channels.
The origin of these event concentrations can be understood as follows.
The concentrations are along the kinematical line of the quasi-free $\bar{K}$ absorption by two residual nucleons, given by
\begin{equation}
\label{eq:m_qf}
m(q) = \sqrt{ 4m_{N}^2 + m_{\bar{K}}^2 + 4m_N \sqrt{m_{\bar{K}}^2+q^2} },
\end{equation}
where $m_N$ and $m_{\bar{K}}$ are the intrinsic masses of the $N$ and $\bar{K}$, respectively.
Equation (\ref{eq:m_qf}) describes the invariant mass of $\bar{K}$ with momentum $q$ interacting with two residual nucleons, thereby representing the kinematics of a quasi-free absorption process in which two residual nucleons absorb an on-shell $\bar{K}$ at rest.
Given that the events are concentrated along this kinematical line, we can interpret this concentration as a manifestation of the quasi-free $\bar{K}$ absorption ($\QF$) process, similar to the non-mesonic $[\Lambda p] + n_{\rm miss}$ channel~\cite{exp/jparc/e15/prc2020}.

If the $\KNN$ quasi-bound state observed in the $[\Lambda p] + n_{\rm miss}$ channel~\cite{exp/jparc/e15/prc2020} decays into the mesonic channels, spectral peaks can be observed in the invariant-mass spectra of each $[\piYN]$ channel below the kaon binding threshold, $m_{\KNN} = m_{\bar{K}}+2m_N$ (indicated by the blue dotted line in the figure).
However, clear peak structures were not observed in these spectra.
The absence of clear peak structures can be understood by considering the density of states in the final state, which is governed by the four-body phase space of $[\piYN] + N_{\rm miss}$. 
The phase space volume below the $m_{\KNN}$ threshold becomes small because the mass region is close to the $\piYN$ threshold, and the four-body phase space opens from the $\piYN$ threshold only in the second order.
As a result, the invariant-mass spectra of $[\piYN]$ in this region are substantially modified because of the limited phase space near the threshold. 
This spectral modification leads to the disappearance of the prominent peak structure.
The density of states in the mesonic final states and the resultant spectral modification are described in greater detail in Sec.~\ref{sec:fit}.

\begin{figure}[ht]
    \centering
    \includegraphics[width=0.8\linewidth]{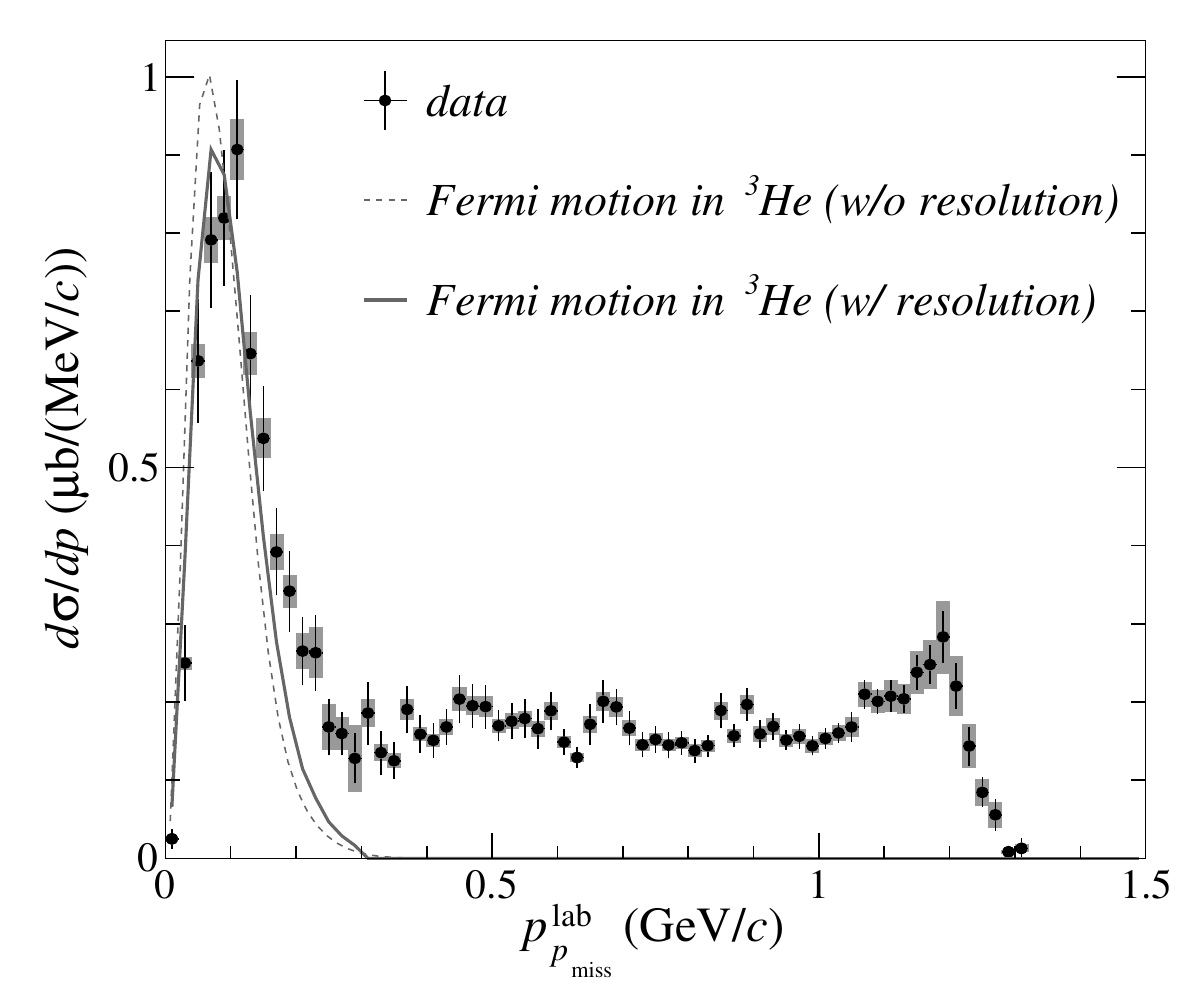}
    \caption{
   Momentum distribution of missing proton ($p_{\rm miss}$) in $(\pi^- \Lambda p + p_{\rm miss})$ channel.
   The gray dotted and solid lines represent the Fermi momentum distribution in ${^3 \rm He}$~\cite{exp/nucleus/jans/prl1982}.
    }
    \label{fig:missing_mom}
\end{figure}

A very strong event concentration is observed at $(m_{[\pi- \Lambda p]},~q_{[\pi^- \Lambda p]}) \sim (2.8\,\GeVcc,\,1\,\GeVc)$ (Fig.~\ref{fig:m_vs_q_cs}(a)) and not in the other $\piYNN$ final states.
Because the width in the $m_{[\pi^- \Lambda p]}$ mass distribution is too narrow to be interpreted as the $\bar{K}$-forward part of the $\QF$ process, the major origin of the events-concentration is the direct two-nucleon $K^-$ absorption (2NA) process by a $pn$ pair, described as \begin{equation}
\label{eq:2na_pimLp}
K^- + (pn) \to \pi^- \Lambda p.
\end{equation}
In this process, the $\bar{K}$ reacts only with a deuteron-like momentum-correlated $pn$ cluster in ${^3 \rm He}$. The other proton in $^3{\rm He}$ acts as a spectator with a Fermi momentum, serving as the missing particle ($p_{\rm miss}$).
We plotted the momentum distribution of $p_{\rm miss}$ in Fig.~\ref{fig:missing_mom} to investigate this interpretation.
As shown in the figure, the momentum distribution of $p_{\rm miss}$ is consistent with our spectator picture, showing a low momentum component, in agreement with the Fermi motion in ${^3 \rm He}$~\cite{exp/nucleus/jans/prl1982}.

By contrast, the same event concentration was not observed in the $[\pi^+ \Lambda n] + n_{\rm miss}$ channels (Fig.~\ref{fig:m_vs_q_cs}(b)).
This result suggests that the $K^-$ beam is less likely to be absorbed by a $pp$ pair than by a $pn$ pair, whose ratio ($pp$/$pn$) is less than $1/10$.
This result is consistent with the absorption of stopped-$\pi^-$ on ${^3 \rm He}~$\cite{exp/nucleus/stopped-pi/prc1995}.

\subsection{$m_{\piY}$ distributions}
\label{sec:results:m_piY}
\begin{figure}[ht]
    \centering
    \includegraphics[width=\linewidth]{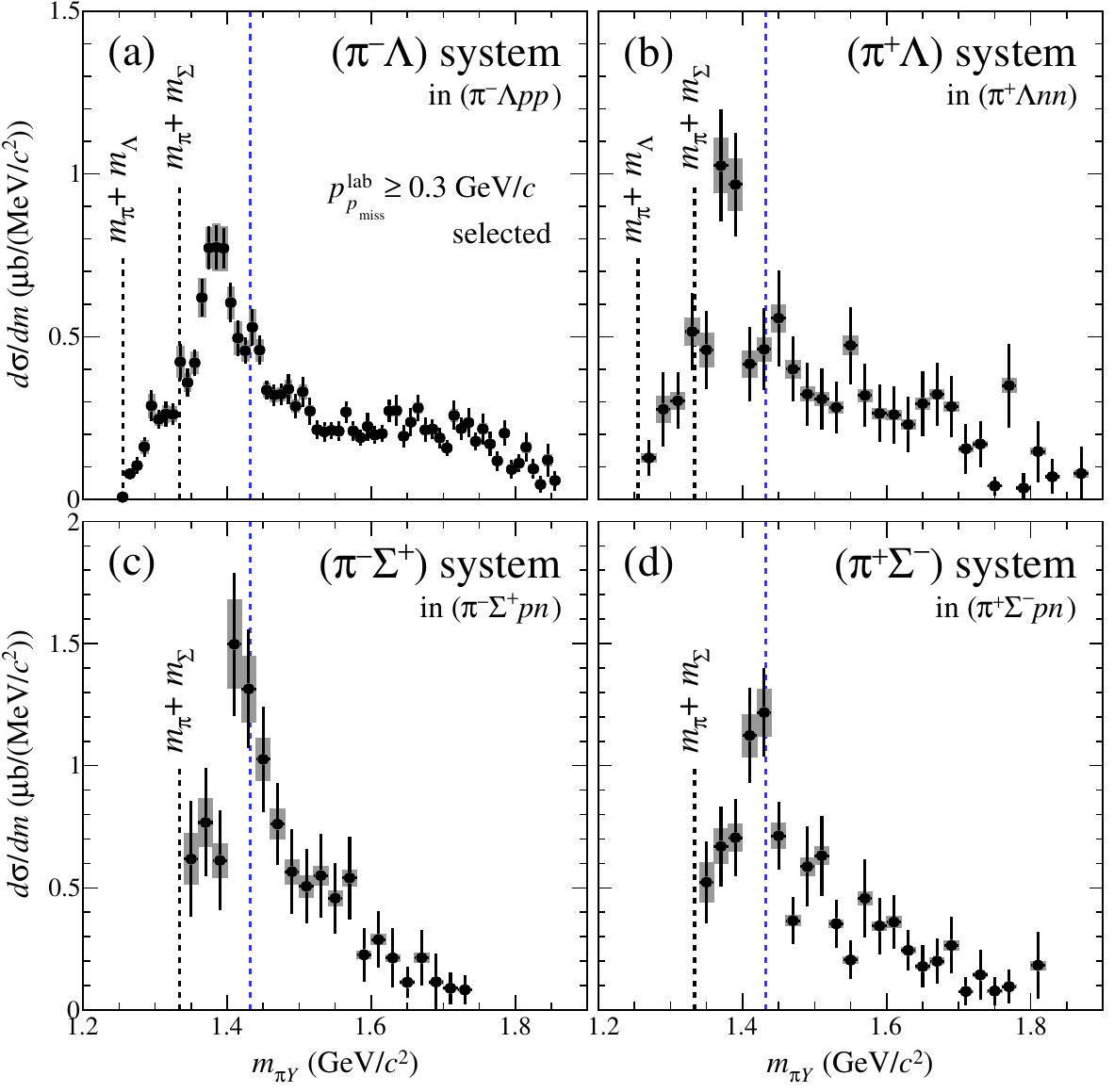}
    \caption{
    Acceptance-corrected $m_{\piY}$ distributions for each $(\piY NN)$ final state. The blue dotted line represents the mass threshold of $m_{\bar{K}} + m_N$ ($m_{\bar{K}N}$).
    }
    \label{fig:m_cs}
\end{figure}

To investigate the role of $Y^*$ resonances in the events coupling to the $Y^*$ poles, we plotted the invariant-mass distribution of the $[\piY]$ subsystem---specifically $d\sigma/dm_{[\piY]}$ (Fig.~\ref{fig:m_cs}). 
For the $[\pi^- \Lambda p] + p_{\rm miss}$ channel, we have excluded events originating from the direct-2NA process by removing events with $p_{p_{\rm miss}}^{\rm lab} < 0.3~\GeVc$. 
This criterion facilitates a more straightforward comparison with other channels.

The figures illustrate that the overall shapes of the $m_{\piY}$ distributions are similar, exhibiting a clear two-body decay of $Y^* \to \piY$ over a broad nonresonant background having similar yields. 
Regarding the $[\pi^\pm \Lambda]$ distributions, the positions of the peaks are consistent with those of $\Sigma(1385)^\pm$. This consistency is attributable to the fixed isospin of $\pi^\pm \Lambda$, resulting in a natural coupling to $\Sigma(1385)^\pm$.
By contrast, the peaks observed in the $[\pi^\mp \Sigma^\pm]$ distributions are slightly greater than $1.4~\GeVcc$. Hence, we infer that these peaks primarily arise from the coupling to the $\Lambda(1405)$ pole. In this channel, the isospin can be either zero or one.
However, the contribution from $\Sigma(1385)^0 \to \pi^\mp \Sigma^\pm$ is expected to be negligible because the decay branching ratio of $\Sigma(1385)$, $Br(\pi\Sigma)/Br(\pi\Lambda)$, is approximately 13\%; if the isospin symmetry holds between $\Sigma(1385)^0$ and $\Sigma(1385)^\pm$, their production yields should be of the same order.
No higher-mass $Y^*$ resonances were observed in the $[\piY]$ distributions.

\begin{figure}[ht]
    \centering
    \includegraphics[width=\linewidth]{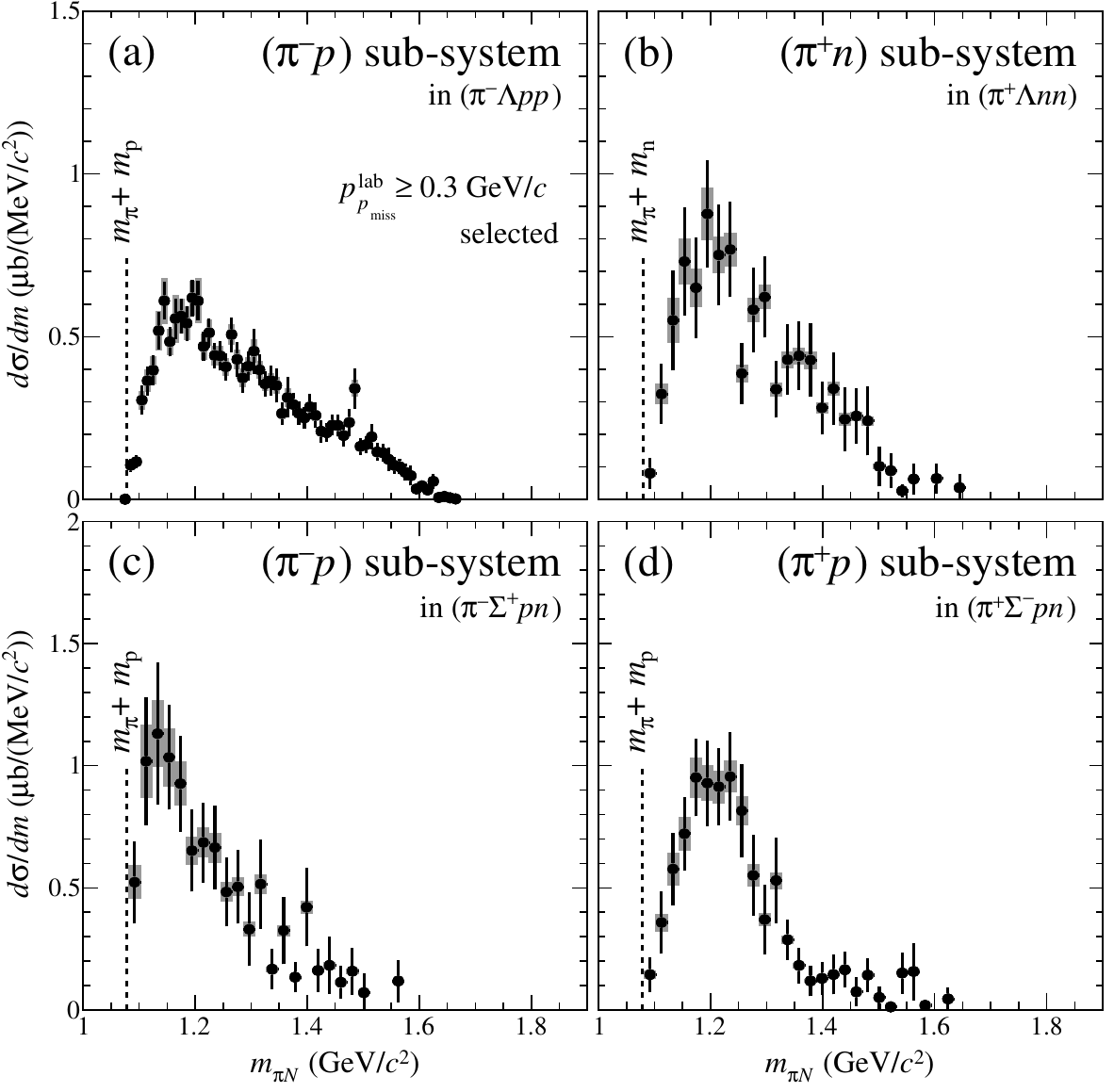}
    \caption{
    Acceptance-corrected $m_{\pi N}$ distributions for each $(\piYNN)$ final state. The black dotted line represents the mass threshold of $m_{\pi} + m_N$.
    }
    \label{fig:m_cs_piN}
\end{figure}

In addition to studying the $[\piY]$ subsystem, we investigated the $[\pi N]$ subsystem to determine if there are any spectral anomalies associated with $\Delta$ or $N^*$ resonances. The considered reactions are
\begin{equation}
K^- + {^3 \rm He} \to Y \Delta N~{\rm or} ~ Y N^* N\to [Y (\pi N)] N_{\rm miss}.
\end{equation}
The resultant $d\sigma/dm_{\pi N}$ spectra are presented in Fig.~\ref{fig:m_cs_piN}.
In contrast to the $m_{\piY}$ distributions, the $m_{\pi N}$ distributions did not show a clear peak or structure.

The results indicate that the dominant processes involve a backward low-momentum $\bar{K}$ in the intermediate state of the reaction via the nucleon knockout reaction $K^-N \to \bar K N'$.
This $\bar{K}$ can efficiently couple with $\Sigma(1385)$ or $\Lambda(1405)$ via $\bar{K} + NN \to Y^*N \to \piYN$.

\section{Discussion}
\label{sec:fit}
As previously described, the overall structures of the mesonic $(m_{[\piYN]},q_{[\piYN]})$ distributions (see Fig.~\ref{fig:m_vs_q_cs}) are similar to the non-mesonic $(m_{[\Lambda p]},q_{[\Lambda p]})$ distribution~\cite{exp/jparc/e15/prc2020}, assuming phase-space suppression of the $\KNN$ peak in mesonic decay channels.
The $m_{[\piY]}$ spectra presented in Fig.~\ref{fig:m_cs} consist of a prominent peak associated with the $\Sigma(1385)$ or $\Lambda(1405)$ resonance, as well as a broad non-resonant contribution.


Therefore, we examined whether the event distribution of the $\piYNN$ final state can be described essentially by the same model functions applied for the $\Lambda p n$ final state, which is composed of $\KNN$ formation and $\QF$, as described in Ref.~\cite{exp/jparc/e15/prc2020}.
Thus, we assumed that the same event distribution for the intermediate $(\bar K + NN)$ system is formed over the $(m_{[\piYN]}, \, q_{[\piYN]})$ kinematical plane, as in the case for the $\Lambda p n$ final state, and that the intermediate $(\bar K + NN)$ decays to $\piYN$.
To allow the $\piY$ subsystem to couple to the $Y^*$ poles in the decay process, we modified the model fitting function $F$ and the phase space accordingly. 
Therefore, the present model fitting function becomes
\begin{equation}
\label{eq:fit}
\begin{split}
F(m,q,m_{\piY}) = \rho(m,q,m_{\piY}) \, \varepsilon(m,q,m_{\piY})  \\  \times 
\sum_j A_j f_j(m,\,q)\big(1+A_{Y^*} \, f_{BW}(m_{\piY}) \big) \\
+ {\rm BG}(m,\,q,\,m_{\piY}),
\end{split}
\end{equation}
where $\rho$ represents the four-body phase-space for $\piYNN$, $\varepsilon$ denotes the acceptance, $f_{BW}$ is a Breit--Wigner function used to describe the $\Sigma(1385)$ or $\Lambda(1405)$ resonance, $A_{Y^*}$ is the coupling strength to the $Y^*$ pole, and $f_j$ and $A_j$ represent the spectral function and strength of each physical process ($j=\KNN$ and $\QF$), respectively.
The parameters $m_{Y^*}$ and $\Gamma_{Y^*}$ of $f_{BW}$ were fixed to the values provided by the Particle Data Group (PDG) in the case of $\Sigma(1385)$~\cite{pdg/rev2022}, whereas we allowed these values to vary in the case of $\Lambda(1405)$.
In the fitting procedure, we ignored any interference between reaction processes.
A resolution function, which smears $f_i$ and $f_{BW}$, was evaluated by the Monte Carlo simulation and convoluted to Eq.~(\ref{eq:fit}).

The $\KNN$ spectral function $f_K$ is defined as
\begin{equation}
\label{eq:fit_knn}
\begin{split}
f_K(m_{\piYN},q_{\piYN}) = \frac{(\Gamma_K/2)^2}{(m_{\piYN}-M_K)^2 + (\Gamma_K/2)^2} \\
\times A_0^K \exp \left( -\frac{q_{\piYN}^2}{Q_K^2} \right),
\end{split}
\end{equation}
where $M_K$, $\Gamma_K$, $Q_K$, and $A_0^K$ are parameters.
The $\QF$ spectral function $f_F$ is a Gaussian distribution along the $M_F(q_{\piYN})$ (Eq.~(\ref{eq:m_qf})) given by
\begin{equation}
\label{eq:fit_qf}
\begin{split}
f_F(m_{\piYN},q_{\piYN}) = \exp \left[ - \frac{ \left( m_{\piYN} - M_F(q_{\piYN}) \right)^2}{\left( \sigma_0 + \sigma_2 , q_{\piYN}^2 \right)^2} \right] \\
\times \left[
A_0^F \exp \left( -\frac{q_{\piYN}^2}{Q_F^2} \right) + A_1^F + A_2^F \exp \left( \frac{m_{\piYN}}{m_0} + \frac{q_{\piYN}}{q_0} \right)
\right],
\end{split}
\end{equation}
where $\sigma_0$, $\sigma_2$, $Q_F$, $m_0$, $q_0$, $A_0^F$, $A_1^F$, and $A_2^F$ are parameters.
In the current fitting, we fixed the parameters of $f_K$ and $f_F$ to those obtained in Ref.~\cite{exp/jparc/e15/prc2020}.

\begin{figure}[ht]
    \centering
    \includegraphics[width=\linewidth]{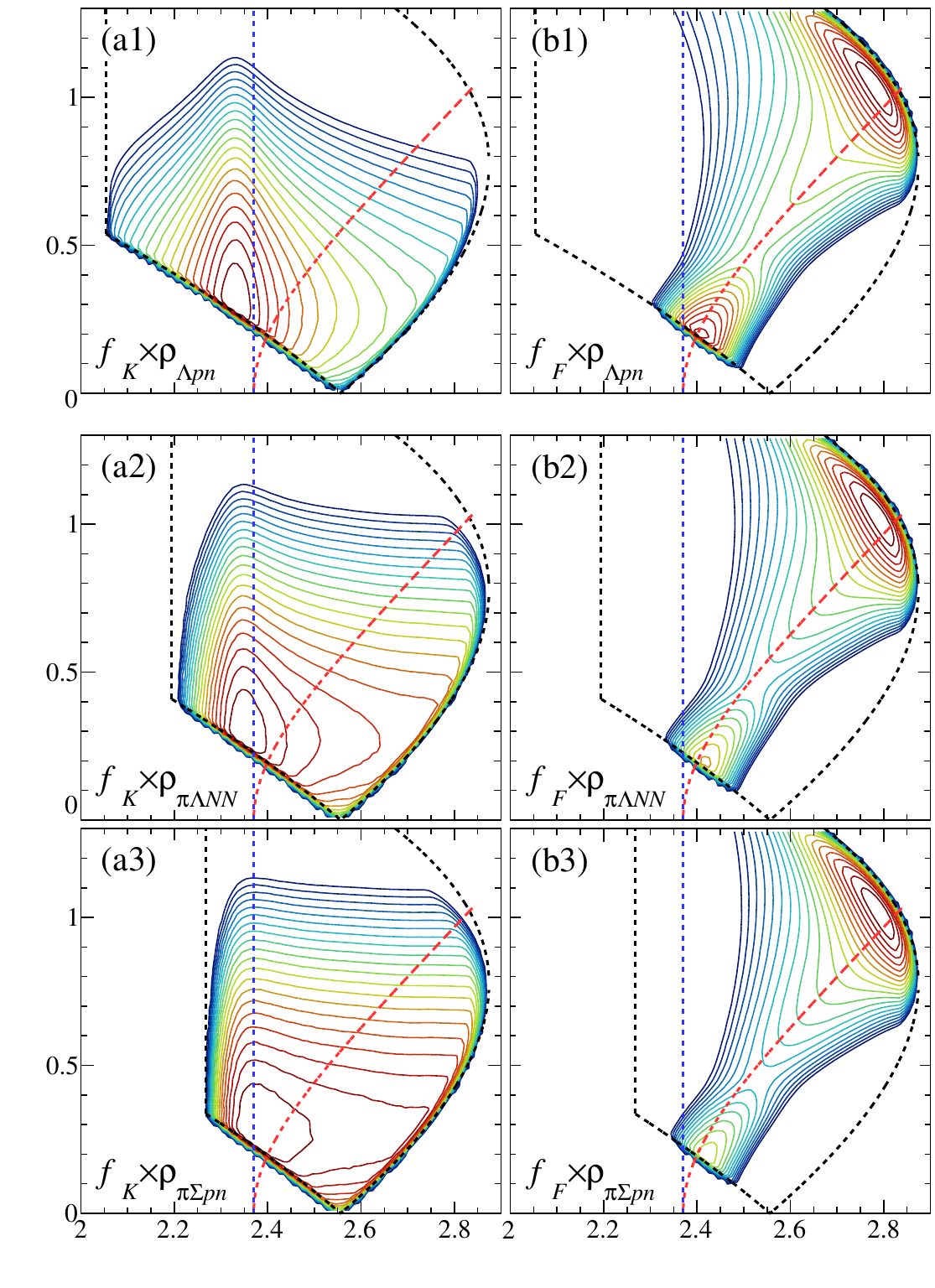}
    \caption{
    Fitting functions for (a1, a2, a3) the $\KNN$ production process and (b1, b2, b3) quasi-free kaon absorption for two-nucleon $QF_{\bar{K}-abs}$ process:
(a1) and (b1) show spectral functions using the three-body phase space of $\Lambda pn$;
(a2) and (b2) correspond to spectral functions using the four-body phase space of $\pi \Lambda NN$;
(a3) and (b3) represent spectral functions using the four-body phase space of $\pi \Sigma pn$.
The $z$ scale is logarithmic.
The black, blue, and red dotted lines correspond to those in Fig.~\ref{fig:m_vs_q_cs}.
    }
    \label{fig:fit_functions}
\end{figure}

To illustrate how the phase space difference can modify spectral functions, we present Fig.~\ref{fig:fit_functions}, which shows two-dimensional model fitting functions for the $\Lambda p$ and $[\piYN]$ spectra. 
The model functions for the $\KNN$ production process in each final state are shown in Fig.~\ref{fig:fit_functions}(a1--a3).
In the $[\Lambda p]+n$ final state (Fig.~\ref{fig:fit_functions}(a1)), the reaction threshold ($m_{\Lambda}+m_{N}$) is substantially smaller than $M_K$ in Eq.~(\ref{eq:fit_knn}), resulting in a nearly flat density of states for $[\Lambda p]$ around the $\KNN$ peak. 
However, in the $[\pi \Lambda N] + N$ (Fig.~\ref{fig:fit_functions}(a2)) and $[\pi^\mp \Sigma^\pm p] + n$ final states (Fig.~\ref{fig:fit_functions}(a3)), the reaction threshold ($m_{\pi} + m_{Y} + m_N$) approaches $M_K$, resulting in substantial suppression of the spectrum because of the diminishing density of states at the lower-mass side of the peak structure; consequently, the maximum of the spectrum appears to shift toward the higher-mass side compared with the $M_K$ value. 
Regarding the $\QF$ process, the spectral modifications in the $[\piYN]+N$ channels are relatively minor compared with those observed in the $\KNN$ case (Fig.~\ref{fig:fit_functions}(b1--b3)). This minor modification in the $\QF$ process is primarily due to the energy region being further separated from the reaction threshold.

This analysis assumes that the detected particles, $\piYN$, originate from the $(\bar{K}+NN)$ intermediate state. However, knocked-out $N'$ in the primary $K^- N \to \bar{K}N'$ reaction can be detected without detecting $N$ from the intermediate state.
In such cases, the $(m,q)$ values for the $[\piY]N'$ system (rather than the $[\piYN]$ system) are distributed widely across the kinematically allowed region. To incorporate this scenario, we generated expected spectral functions through simulations for each relevant process and summed them to account for the background, ${\rm BG}(m,~q,~m_{\piY})$.
There is background contamination resulting from the misidentification of the final state, as described in Sec.~\ref{sec:analysis:event-selection}.
In the present fitting, we omitted this background; however, we considered its effect when evaluating the cross-section systematic error.

\begin{figure}[h]
\centering
\includegraphics[width=\linewidth]{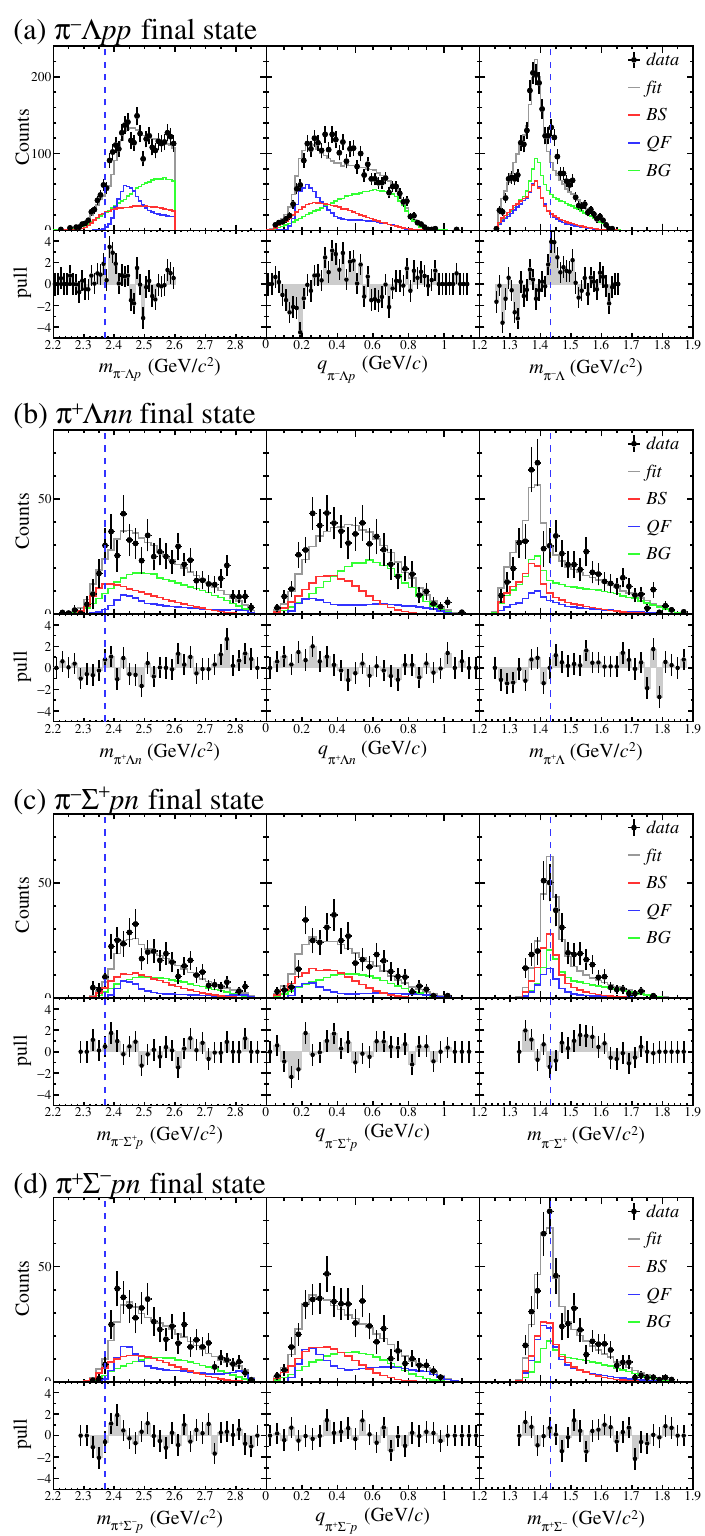}
\caption{
 Fitted distributions and results of fitting for channels (a) $[\pi^- \Lambda p] + p_{\rm miss}$, (b) $[\pi^+ \Lambda n] + n_{\rm miss}$, (c) $[\pi^- \Sigma^+ p] + n_{\rm miss}$, and (d) $[\pi^+ \Sigma^- p] + n_{\rm miss}$.
 ``BS'', ``QF'', and ``BG'' denote $\KNN$ production, $\QF$, and background processes, respectively.
}
\label{fig:fit_result}
\end{figure}

Because of the limited statistics for the limited acceptance on the three-dimensional kinematical space of $(m_{[\piYN]}, \,q_{[\piYN]}, \, m_{[\piY]})$, neither a multi-dimensional fitting nor the application of the coupled-channel-based sophisticated fitting function (to account for the threshold effects) is feasible in this analysis. 
Instead, we performed a simultaneous fit of three one-dimensional spectra along the $m_{[\piYN]}$, $q_{[\piYN]}$, and $m_{[\piY]}$ axes.
The two charged modes, $[\pi^\mp \Sigma^\pm p] + n_{\rm miss}$, were fitted simultaneously using common values of $M_{\Lambda(1405)}$ and $\Gamma_{\Lambda(1405)}$.
The fitting results are shown in Fig.~\ref{fig:fit_result}, and the obtained parameters are summarized in Table~\ref{tab:fit_parameters}.
The chi-square values and the number of degrees of freedom for $[\pi^- \Lambda p] + p_{\rm miss}$, $[\pi^+ \Lambda n] + n_{\rm miss}$, and $[\pi^\mp \Sigma^\pm p] + n_{\rm miss}$ were $\chi^2/NDF = 275.5/118$, $76.3/85$, and $114.9/136$, respectively.
The figures demonstrate that the proposed model functions provide a good overall data description.


\begin{table}[h]
\centering
\renewcommand{\arraystretch}{1.3}
\caption{Parameters obtained by fitting.}
\label{tab:fit_parameters}
\begin{tabularx}{\linewidth}{RR}
\hline
\hline
\multicolumn{2}{c}{$A_i$ for $(\pi^- \Lambda p)$ channel} \\
$A_{K}$                      & $1140.5 \pm 112.2$      \\   
$A_{F}$                      & $1800.6 \pm 77.0$      \\   
$A_{Y^*}$                    & $1.4  \pm 0.1$        \\   
\hline
\multicolumn{2}{c}{$A_i$ for $(\pi^+ \Lambda n)$ channel} \\
$A_{K}$                      & $2707.7 \pm 467.3$      \\   
$A_{F}$                      & $1557.5 \pm 114.5$      \\   
$A_{Y^*}$                    & $1.2   \pm 0.4$        \\   
\hline
\multicolumn{2}{c}{$A_i$ for $(\pi^- \Sigma^+ p)$ channel} \\
$A_{K}$                      & $2010.6 \pm 57.3$      \\   
$A_{F}$                      & $1167.9 \pm 138.8$      \\   
$A_{Y^*}$                    & $4.4 \pm 0.6$        \\   
\hline
\multicolumn{2}{c}{$A_i$ for $(\pi^+ \Sigma^- p)$ channel} \\
$A_{K}$                      & $2159.7 \pm 150.8$      \\   
$A_{F}$                      & $2583.9 \pm 108.1$      \\   
$A_{Y^*}$                    & $1.8 \pm 0.2$        \\   
\hline
\multicolumn{2}{c}{parameters for $\Lambda(1405)$} \\
$M_{\Lambda(1405)}$          & $1432 \pm 5 ~\MeVcc$       \\   
$\Gamma_{\Lambda(1405)}$     & $49 \pm 3 ~\MeVcc$       \\   
\hline
\hline
\end{tabularx}
\end{table}

In Figs.~\ref{fig:fit_result}(c) and \ref{fig:fit_result}(d), the $\Lambda(1405)$ contribution in the $m_{\pi^\mp \Sigma^\pm}$ distributions is well reproduced by a simple Breit--Wigner function. The obtained mass and width of $\Lambda(1405)$ are $1432\pm5~\MeVcc$ and $49\pm3~\MeVcc$, respectively. These values are close to the pole position deduced from the $d(K^-,n)\pi\Sigma$ reactions, $(m,\Gamma) = (1418~\MeVcc,52~\MeVcc)$\cite{exp/jparc/e31/plb2023}, and to the theoretically predicted pole positions based on chiral unitary approaches, $(1421-1434~\MeVcc,20-52~\MeVcc)$~\cite{theor/lambda1405/ikeda/plb2011,theor/lambda1405/ikeda/npa2012,theor/lambda1405/guo/prc2013,theor/lambda1405/mai/epja2015,theor/lambda1405/liu/prd2017,theor/lambda1405/lu/prl2023}.

\begin{figure*}[ht]
\centering
\includegraphics[width=\linewidth]{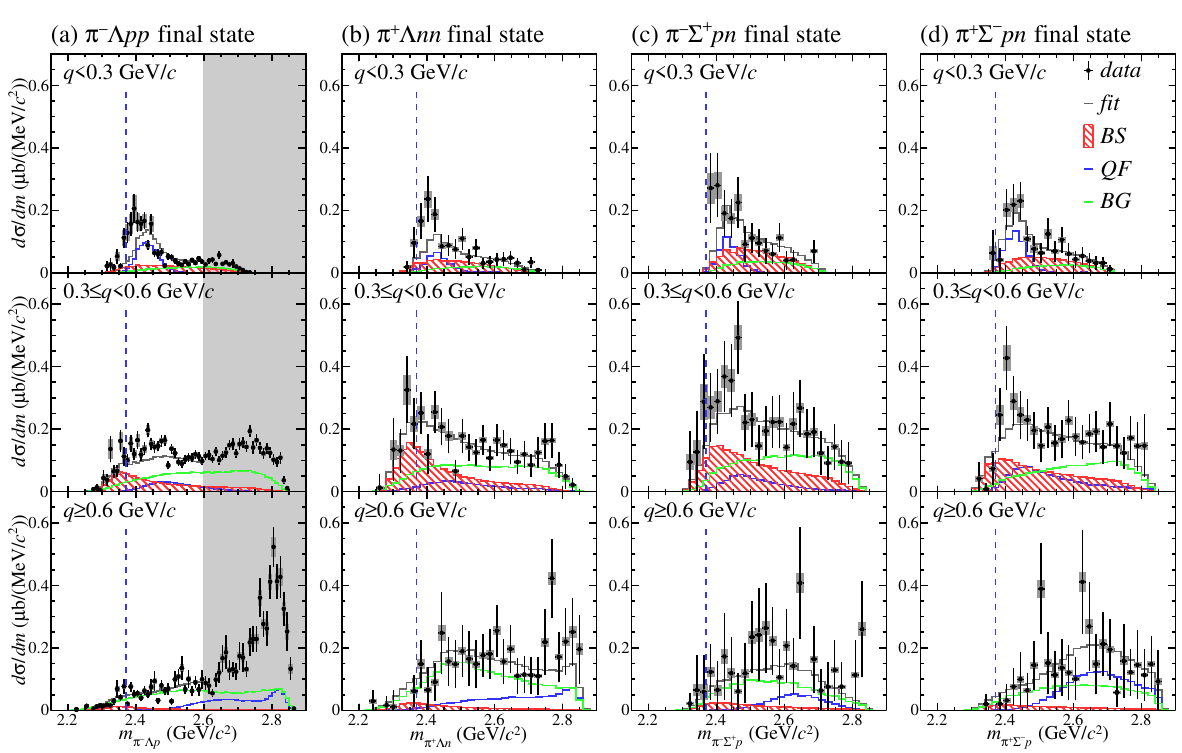}
\caption{
Differential cross-sections $d\sigma/dm_{\piYN}$ are shown in each $q$ interval for (a) $[\pi^- \Lambda p] + p_{\rm miss}$, (b) $[\pi^+ \Lambda n] + n_{\rm miss}$, (c) $[\pi^- \Sigma^+ p] + n_{\rm miss}$, and (d) $[\pi^+ \Sigma^- p] + n_{\rm miss}$ channels.
The colored histograms represent the fitting results obtained from the analysis.
In panel (a), the gray hatched box highlights a region outside the fitting range to exclude the direct 2NA process.
The blue dotted lines indicate the $m_{\KNN}$ threshold.
}
\label{fig:fit_result_cs_piYN}
\end{figure*}

\begin{figure*}[ht]
\centering
\includegraphics[width=\linewidth]{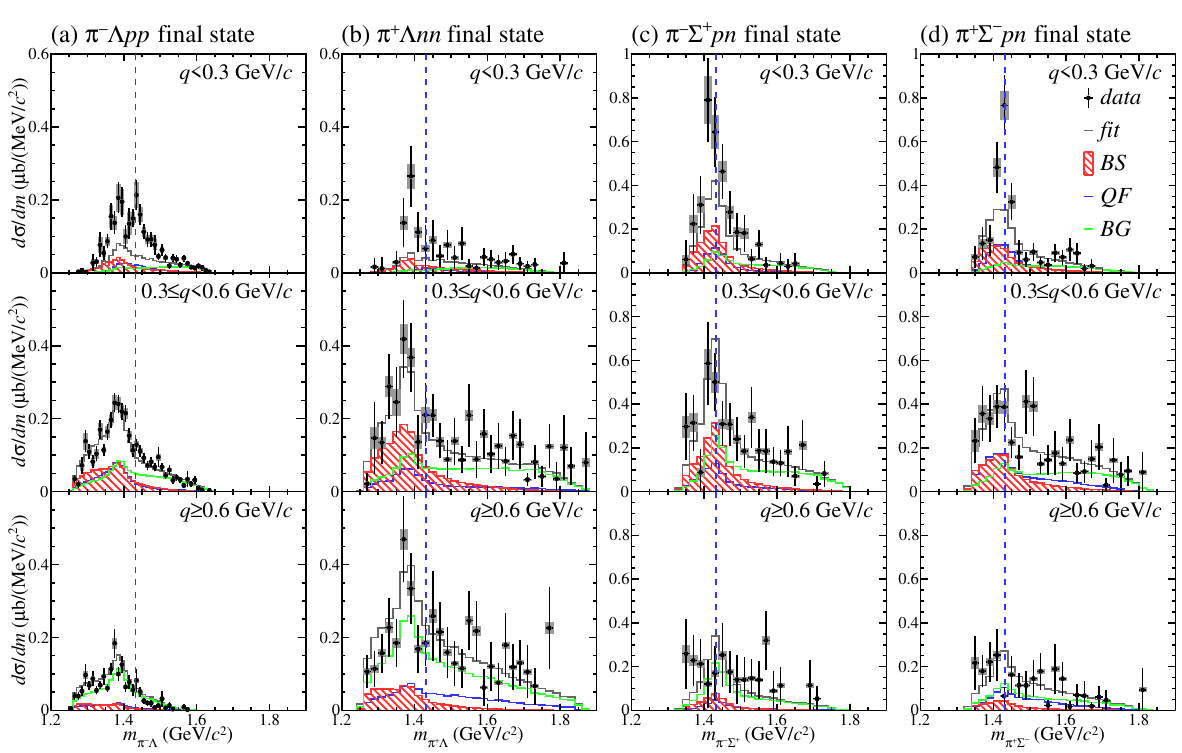}
\caption{
Differential cross-sections $d\sigma/dm_{\piY}$ are shown in each $q$ interval for (a) $[\pi^- \Lambda p] + p_{\rm miss}$, (b) $[\pi^+ \Lambda n] + n_{\rm miss}$, (c) $[\pi^- \Sigma^+ p] + n_{\rm miss}$, and (d) $[\pi^+ \Sigma^- p] + n_{\rm miss}$ channels.
The colored histograms represent the fitting results obtained from the analysis.
In the case of (a), the requirements of $m_{\pi^- \Lambda p}<2.6~{\rm GeV}/c^2$ and $p_{\rm miss}>0.3~\GeVc$ are imposed to exclude the direct 2NA process.
The blue dotted lines indicate the $m_{\KN}$ threshold.
}
\label{fig:fit_result_cs_piY}
\end{figure*}

We examined the validity of the global fit using $q_{[\piYN]}$-sliced mass spectra, as shown in Figs.~\ref{fig:fit_result_cs_piYN} and \ref{fig:fit_result_cs_piY}.
The model functions successfully reproduced the global structure of each sliced distribution except for the $m_{[\piY]}$ spectra near the $\KN$ threshold in the low-momentum-transfer region of $q<0.3~{\rm GeV}/c$, as shown in Fig.\ref{fig:fit_result_cs_piY}.
This discrepancy might arise from the effect of the neglected channel coupling with the $\KN$ channel ({\it i.e.}, the $\bar{K}N$ threshold cusp in the $I_{\bar{K}N}=1$ channel recently reported by the Belle collaboration~\cite{exp/belle/ma/prl2023}).



The total cross-sections of $\KNN$ decaying into each mesonic decay channel, $\sigma_{\KNN}^{tot} \, Br(\piYN)$, were evaluated by integrating the $\KNN$ fitting function over the entire kinematical region including above the $\bar{K}$ binding threshold, $m_{\bar{K}}+2m_N$, as summarized in Table~\ref{tab:cs_branch}.
We also tabulated the branch below the $m_{\bar{K}}+2m_N$ threshold to exclude the energy region where the spectral shape of the fitting functions would become less reliable because of threshold effects.
The values for non-mesonic decay modes, $\KNN_{I_3=+1/2} \to \Lambda p$ and $\Sigma^0 p$, were taken from Ref.~\cite{exp/jparc/e15/prc2020}.
The values for the $\KNN \to \Sigma^+ n$ and $\KNN \to \pi^0 \Lambda p$ channels were evaluated as $Br(\Sigma^+ n) = 2 Br(\Sigma^0 p)$ and $Br(\pi^0 \Lambda p) = 1/2Br(\pi^+ \Lambda n)$, respectively, on the basis of the isospin symmetry.
The total non-mesonic decay branch was then evaluated as the sum of the three non-mesonic channels, the errors of which are the linear sum of these three channels because the errors are correlated.  

We found that the ratio of mesonic to non-mesonic decay branches is $\mathcal{O}(10)$ if we integrate all mass regions.
The mesonic decay branch is still larger than the non-mesonic one, even if we integrate below the $\bar{K}$ binding threshold.
These results suggest that the mesonic decay mode is the dominant decay branch of the $\KNN$, as expected from theoretical considerations.

The decay branches $Br(\pi^+ \Lambda n)$ and $Br(\pi^\mp \Sigma^\pm p)$ were similar in magnitude in both integration ranges.
The results suggest that the mesonic decay modes coupled to the $I_{\KN}=1$ channel play a substantial role in the $\KNN$ decay process.
In addition, this coupling could explain why the decay width of the $\KNN$ is much broader than that of the $\Lambda(1405)$ resonance: the mesonic decay modes with $I_{\KN}=1$ contribute to the broadening of the $\KNN$ decay width. \begin{table}[h]
\centering
\renewcommand{\arraystretch}{1.3}
\caption{
Cross-sections of each decay channel for $\KNN$.
The first and second errors are statistical and systematic ones, respectively.
The $\bar{K}$ binding thresholds for $I_{3} = \pm 1/2$ were considered as $m_{K^-}+2m_p$ and $m_{\bar{K}^0}+2m_n$, respectively.
We estimated decay branches for $\Sigma^+ n$ and $\pi^0 \Lambda p$ channels by assuming isospin symmetry.
}
\label{tab:cs_branch}
\begin{tabularx}{\linewidth}{RRR}
\hline
\hline
 & \multicolumn{2}{c}{$\sigma_{\KNN}^{tot} \, Br~(\mu {\rm b})$} \\
Decay channel & All regions & Below the $\bar{K}$ binding threshold \\
\hline
 \multicolumn{3}{l}{$\KNN_{I_3=+1/2}$} \\
 $\Lambda p$   & $9.3 \pm 0.8 ^{+1.4}_{-1.0}$~\cite{exp/jparc/e15/prc2020}  & $5.5 \pm0.5^{+0.8}_{-0.6}$ \\   
 $\Sigma^0 p$  & $5.3 \pm 0.4 ^{+0.8}_{-0.6}$~\cite{exp/jparc/e15/prc2020}  & $3.1 \pm0.2^{+0.5}_{-0.4}$  \\   
 $\Sigma^+ n^{(*)}$ ($= \Sigma^0 p \times 2$) & $10.6 \pm 0.8 ^{+1.6}_{-1.2}$  & $6.2 \pm0.4^{+1.0}_{-0.8}$    \\   
 total non-mesonic & $25.2 \pm 2.0 ^{+3.8}_{-2.8}$ & $14.8 \pm1.1^{+2.3}_{-1.8}$ \\
$\pi^0 \Lambda p^{(*)}$ ($= \pi^+ \Lambda n \times 1/2$)   & $31 \pm 5.5 \pm 4.5$    & $7.8 \pm 1.4 \pm 1.1$  \\   
$\pi^0 \Sigma^0 p$  & NA    & NA \\   
$\pi^- \Sigma^+ p$  & $110 \pm 8 \pm8$   & $9.4\pm0.4\pm0.7$   \\ 
$\pi^+ \Sigma^- p$  & $38 \pm3 \pm3$     & $3.2\pm0.2\pm0.2$ \\   
$\pi^+ \Lambda  n$  & $62 \pm11 \pm9$    & $15.5\pm2.7\pm2.1$ \\   
$\pi^+ \Sigma^0 n$  & NA & NA      \\   
$\pi^0 \Sigma^+ n$  & NA & NA      \\   
total mesonic  & $> 241 \pm 20 \pm 17$ & $> 37.9\pm4.1 \pm3.3$\\
\hline
 \multicolumn{3}{l}{$\KNN_{I_3=-1/2}$} \\
 $\pi^- \Lambda p$   & $29 \pm3 \pm3$ & $7.2 \pm0.6 \pm0.7$  \\   
\hline
\hline
\end{tabularx}
\end{table}

\section{Summary}
We conducted measurements of the $K^- + {^3 \rm He}$ reactions resulting in mesonic final states---namely, $\pi^\mp \Sigma^\pm p + n'$, $\pi^+ \Lambda n + n'$, and $\pi^- \Lambda p + p'$---using an incident $K^-$ momentum of $1~\GeVc$.
To investigate the mesonic decay of the $\KNN$ state, we measured the two-dimensional distributions of the invariant mass and momentum transfer of the $\piYN$ systems.
The observed two-dimensional distributions of the $\piYN$ systems were well described by the model functions, which included contributions from $\KNN$ production and quasi-free $\bar{K}$ absorption processes as a result of the reactions given in Eqs.~(\ref{eq:knockout}) and (\ref{eq:KNN_to_piYN}). 
These model functions are essentially the same as those introduced in Ref.~\cite{exp/jparc/e15/prc2020} to explain the $K^- + {^3 \rm He} \to \Lambda p + n'$ reaction by incorporating appropriate phase-space distributions.

The fitting results suggest that the mesonic decay is the dominant decay branch of the $\KNN$. 
The results also suggest that the branching ratio of $\KNN \to \pi^+ \Lambda n$ is similar to that of $\KNN \to \pi^\mp \Sigma^\pm p$, indicating that the mesonic $\bar{K}N$ absorption process in the $I_{\bar{K}N}=1$ channel plays an important role, similar to the process in the $I_{\KN}=0$ channel.The significant fraction of decay into the $I_{\bar{K}N}=1$ channel---specifically, the $\pi \Lambda N$ decay---would be a major factor contributing to the broader decay width of the $\KNN$ resonance compared with that of the $\Lambda(1405)$.
The presence of the $I_{\KN}=1$ cusp in the $\piY$ subsystem (Fig.~\ref{fig:fit_result_cs_piY}) suggests that the real part of the $I_{\KN}=1$ interaction is also attractive, although it is not sufficiently strong to form a bound state in this channel.

The branching ratio for the $\KNN$ between the non-mesonic and mesonic channels provides valuable information regarding the size of the system because the strength of mesonic decay is directly proportional to the nucleon density. By contrast, the strength of non-mesonic decay is proportional to the square of the nucleon density because it requires simultaneous interaction of the antikaon with two nucleons.
On the basis of the present study, the $Br(YN)/Br(\piYN)$ ratio is estimated to be approximately 1/10, as obtained via integration over all mass regions. However, we note that the ratio becomes smaller if the integration range is limited to the $\bar{K}$ binding threshold.
A more precise analysis, especially one using a more realistic model function, is necessary to compare the experimental results with the theoretical calculations.
This analysis would provide more information about the size of the system.

Further study of the multi-nucleon absorption of the $K^-$ beam could provide valuable insights into the clustering mechanism within the nucleus. 
The data show that the kaon's direct two-nucleon absorption process is predominantly observed in the "$pn$" pair. 
Conversely, almost no reaction is observed in the "$pp$" pair. 
This difference can intuitively indicate a stronger coupling or clustering of the ($pn$) pair compared with that of the ($pp$) pair in ${^3 \rm He}$.

It is crucial to conduct new experiments with enhanced acceptance coverage and improved neutron detection efficiency to understand these issues better and investigate the production mechanism for kaonic nuclei. 
These advancements will enable us to gather more comprehensive data and will provide valuable insights into the phenomena under investigation. By increasing the experimental capabilities, we can further explore the properties and behavior of kaonic nuclei, shed light on their production mechanisms, and contribute to a more thorough understanding of these fascinating systems~\cite{book/iwasaki_knucl_exp}.

\begin{acknowledgements}
The authors are grateful to the staff members of J-PARC/KEK for their extensive efforts, especially in the stable operation of the facility. We are also grateful for the contributions of Professors D.~Jido, T.~Sekihara, Y.~Akaishi, A.~Dote, T.~Harada, O.~Morimatsu, and J.~Yamagata and Dr.~K.~Suzuki. This work is partly supported by MEXT Grants-in-Aid 14102005, 17070007,  24105003, 26287057, 26800158, 17K05481, 18H05402, 20K04006, 21K13952, 22H04917, 22H04940, and 24H00029. Part of this work is supported by the EU STRONG-2020 project
(Grant Agreement No.~824093) and by the EXOTICA project of the Minstero degli Affari Esteri e della Cooperazione Internazionale, PO22MO03.
\end{acknowledgements}

\bibliography{main}

\end{document}